\documentclass[%
reprint,
superscriptaddress,
nofootinbib,
amsmath,amssymb,
aps,
prd,
]{revtex4-2}
\usepackage[
    colorlinks=true,
    linkcolor=WildStrawberry,
    citecolor=RoyalBlue,
    urlcolor=RedViolet,
    ]{hyperref}
\usepackage{cleveref}

\usepackage{graphicx}%
\usepackage{amsfonts}%
\usepackage{amsthm}%
\usepackage{mathrsfs}%
\usepackage{braket}%
\usepackage{dsfont}%
\usepackage{mathtools}%
\usepackage[dvipsnames]{xcolor}%
\usepackage{etoolbox}%

\theoremstyle{remark}%
\theoremstyle{definition}%
\theoremstyle{plain}

\newcommand\I{\mathrm{i}}
\newcommand\E{\mathrm{e}}
\newcommand\D{\mathrm{d}}

\newcommand\R{\mathbb{R}}

\newcommand\N{\mathbb{N}}
\newcommand\Z{\mathbb{Z}}

\begin{document}

% Use the \preprint command to place your local institutional report
% number in the upper righthand corner of the title page in preprint mode.
% Multiple \preprint commands are allowed.
% Use the 'preprintnumbers' class option to override journal defaults
% to display numbers if necessary
%\preprint{}

%Title of paper
\title{Time Evolution of Relativistic Quantum Fields in Spatial Subregions}

% repeat the \author .. \affiliation  etc. as needed
% \email, \thanks, \homepage, \altaffiliation all apply to the current
% author. Explanatory text should go in the []'s, actual e-mail
% address or url should go in the {}'s for \email and \homepage.
% Please use the appropriate macro foreach each type of information

% \affiliation command applies to all authors since the last
% \affiliation command. The \affiliation command should follow the
% other information
% \affiliation can be followed by \email, \homepage, \thanks as well.
\author{Markus Schr\"{o}fl}
\email[]{schroefl@thphys.uni-heidelberg.de}
\affiliation{Institut f\"{u}r Theoretische Physik, Universit\"{a}t Heidelberg, Philosophenweg 16, 69120 Heidelberg, Germany}
\affiliation{Theoretisch-Physikalisches Institut, Friedrich-Schiller-Universität Jena, Max-Wien-Platz 1, 07743 Jena, Germany}

\author{Stefan Floerchinger}
\email[]{stefan.floerchinger@uni-jena.de}
%\homepage[]{Your web page}
%\thanks{}
%\altaffiliation{}
\affiliation{Theoretisch-Physikalisches Institut, Friedrich-Schiller-Universität Jena, Max-Wien-Platz 1, 07743 Jena, Germany}

%Collaboration name if desired (requires use of superscriptaddress
%option in \documentclass). \noaffiliation is required (may also be
%used with the \author command).
%\collaboration can be followed by \email, \homepage, \thanks as well.
%\collaboration{}
%\noaffiliation

\date{\today}

\begin{abstract}
    We study the time evolution of a state of a relativistic quantum field theory restricted to a spatial subregion $\Omega$. More precisely, we use the Feynman-Vernon influence functional formalism to describe the dynamics of the field theory in the interior of $\Omega$ arising after integrating out the degrees of freedom in the exterior. We show how the influence of the environment gets encoded in a boundary term. Furthermore, we derive a stochastic equation of motion for the field expectation value in the interior. We find that the boundary conditions obtained in this way are energy non-conserving and non-local in space and time. Our results find applications in understanding the emergence of local thermalization in relativistic quantum field theories and the relationship between quantum field theory and relativistic fluid dynamics.
\end{abstract}

% insert suggested keywords - APS authors don't need to do this
%\keywords{}

%\maketitle must follow title, authors, abstract, and keywords
\maketitle

%\tableofcontents

% body of paper here - Use proper section commands
% References should be done using the \cite, \ref, and \label commands
\section{Introduction}
% Put \label in argument of \section for cross-referencing
%\section{\label{}}

The time evolution of an isolated quantum system is unitary and therefore its entropy is constant in time. For example, a pure state of a closed quantum system remains pure during its time evolution. Nevertheless, an isolated many-body quantum system in a non-equilibrium state can thermalize under such unitary dynamics \cite{Deutsch1991,Srednicki1994,Rigol2008}. This is especially the case if one considers only the expectation values of observables supported on a (small) subsystem \cite{Eisert2015,Kaufman2016}. A driving principle for the thermalization of many-body quantum systems is thought to be the generation of entanglement between subsystems \cite{Kaufman2016,Popescu2006,Abanin2019}, which can also be measured experimentally \cite{Islam2015}.

Thermalization caused by entanglement generation may also be important in high-energy experiments \cite{Berges2018a,Berges2018b,Berges2018c}. Moreover, a detailed understanding of the ``local'' subsystem dynamics and the related concepts of entanglement generation and entropy increase may be crucial for understanding the relationship between quantum field theory and relativistic fluid dynamics \cite{Dowling2020}. Relativistic fluid dynamics \cite{Landau2013,Israel1979,Kovtun2012} provides a powerful phenomenological description of the dynamics of quantum fields, for example in the context of heavy ion collisions \cite{Teaney2010,Heinz2013,Busza2018}. One attempt to understand the relationship between quantum field theory and relativistic fluid dynamics is based on the concept of local thermal equilibrium. It is assumed that spatial subsystems, which are small compared to the scale on which typical experiments are conducted, are \emph{open} quantum systems, i.e., they are not isolated from their environment. Therefore, such a local time evolution is \emph{non-unitary}, since information can be exchanged with the environment. In contrast to the unitary time evolution of closed systems, entropy is no longer constant under non-unitary time evolution, and entropy can increase locally. This is thought to be one of the reason for the emergence of local thermal equilibrium and thus the applicability of relativistic fluid dynamics to quantum field theories.

In this work, we aim to gain insight into the emergence of local thermalization of relativistic quantum field theories by studying the structure of the local dynamics in a relativistic field theory. As a toy model, we consider a massive scalar field on $d = D+1$ dimensional Minkowski spacetime. We are interested in the dynamics of the field theory in a spatial subregion $\Omega$. Defining the field within this region as the ``system'' and the field outside the region as the ``environment'', the differential operator in the action of the field theory induces a linear system-environment coupling similar to the Caldeira-Leggett model \cite{Caldeira1983a,Caldeira1983b,Caldeira1985}. This coupling leads to open dynamics of the system, i.e. the time evolution of the reduced state of the system is non-unitary, since information can dissipate between the interior and exterior regions. As described above, such a non-unitary time evolution is a necessary condition for an increase of entropy and thus for local thermalization.

Besides the presence of a system-environment interaction, the initial state of a field theory plays a crucial role in determining the system's dynamics. Therefore, it is important to consider the implications of the initial state when performing calculations. Although an uncorrelated initial state is typically assumed for practical reasons, this assumption is difficult to justify in the case of relativistic field theories. Thermal states, including the vacuum, are correlated across space. Even when classical correlations are absent (like in the vacuum state), it is well-known that states of relativistic quantum field theories are highly (and in some sense even maximally) entangled \cite{Witten2018,Hollands2018}. As a result, our analysis must account for the possibility of initial state correlations, which further complicates the problem.

The main objective of this work is to derive the effective dynamics of a local field theory in a model where the environment integrals are exactly solvable. In particular, we assume the initial state of the environment to be Gaussian and the dynamics of the environment to be linear. Furthermore, we derive a stochastic equation of motion for the field expectation values in the interior. Due to the local nature of the theory, the stochastic parts of the dynamics are encoded in the spatial boundary conditions of the differential equation. We show that for the initial states considered in this work, the boundary conditions derived from the effective dynamics are reminiscent of so-called ``free'' boundary conditions as described in \cite{Guerra1975a,Guerra1975b,Guerra1976}. Finally, we extend the discussion to a polynomially self-interacting field and show that in this case the effective dynamics of the reduced theory are again encoded in an effective boundary term.

In most of this work we consider a lattice regularized theory. This allows us to make the system-environment decomposition more transparent and to derive the effective dynamics of the system in a controlled way. However, we will introduce a suggestive notation for summation and difference quotients, which will allow the reader to straightforwardly read off the corresponding continuum expressions. We explicitly state whenever we work with the continuum theory. Only in Section \ref{sec:Dynamics_of_Field_Expectation_Values} do we work directly in the continuum, since the discussion of the partial differential equation for the field expectation value is more natural in this setting. Since it would be beyond the scope of this work to rigorously derive the continuum limit of the lattice theory, we will not go into the details of the mathematical subtleties involved in this procedure. However, at least for the free theory with non-pathological choices of the interior region (i.e. the interior and its boundary are assumed to be sufficiently regular in the continuum theory), we expect that taking the continuum limit should in principle not pose any problems. For the interacting theory discussed in Section \ref{sec:Interacting_Theories}, one is faced with the usual technicalities of renormalization, the details of which when working in configuration space instead of Fourier space and on a finite volume would be interesting in themselves.

\emph{The remainder of this paper is organized as follows.} In Section \ref{sec:Lattice_Model}, we introduce the main ideas of this work in the context of a lattice regularized scalar field theory. We define the lattice model, introduce the lattice Laplacian and discuss the decomposition of the action of the field theory into interior, exterior and boundary parts. We test our results by considering a reduced state of a Euclidean field theory on the lattice, where we recover a field theory with free boundary conditions as discussed in \cite{Guerra1975a,Guerra1975b,Guerra1976}. In Section \ref{sec:Local_Dynamics}, we derive the effective local dynamics of a free massive scalar field theory. We show that the effects of the environment on the system are encoded in an effective boundary term. Furthermore, we derive an equation of motion for the field expectation values. The implications of self-interactions on the effective local dynamics are discussed in Section \ref{sec:Interacting_Theories}. In Section \ref{sec:Conclusion}, we summarize our results and discuss possible future directions.

\emph{Notation and Definitions.} The time derivative of a function $f$ is denoted by a dot, i.e., $\dot{f} \coloneqq \partial f / \partial t$. A superscript $^\mathsf{T}$ denotes the transpose of a matrix or vector. Throughout this paper, we work in natural units, i.e., $\hbar = c = k_\mathrm{B} = 1$.

\section{Lattice Model}\label{sec:Lattice_Model}

In this Section we consider a lattice regularized model of a relativistic scalar field theory. More precisely, we discretize space into a regular lattice and compactify space into a torus, yielding a system with finitely many degrees of freedom with significantly fewer subtleties than the continuum theory. In particular, the splitting of the model into a ``system'' (the degrees of freedom inside some region $\Lambda$) and an ``environment'' (the degrees of freedom in the complement of $\Lambda$) is more transparent if we consider a lattice theory. In Section \ref{sec:Lattice_Laplacian}, we define the lattice Laplacian and its associated quadratic form and discuss the decomposition of the latter into interior, exterior and boundary parts. We then test the results from Section \ref{sec:Lattice_Laplacian} in Section \ref{sec:Euclidean_Field_Theory} by considering a reduced state of a Euclidean field theory on the lattice and show that we recover a Euclidean field theory with free boundary conditions as discussed in \cite{Guerra1975a,Guerra1975b,Guerra1976}. Finally, we briefly introduce a \emph{quantum} lattice field theory in Section \ref{sec:Quantum_Lattice_System} and use the results from \ref{sec:Lattice_Laplacian} to split the Hamiltonian of the theory into system, environment and boundary parts.

\subsection{Lattice Regularization and Lattice Laplacian}\label{sec:Lattice_Laplacian}

\begin{figure}[t]
    \centering
    \includegraphics[width=0.35\textwidth]{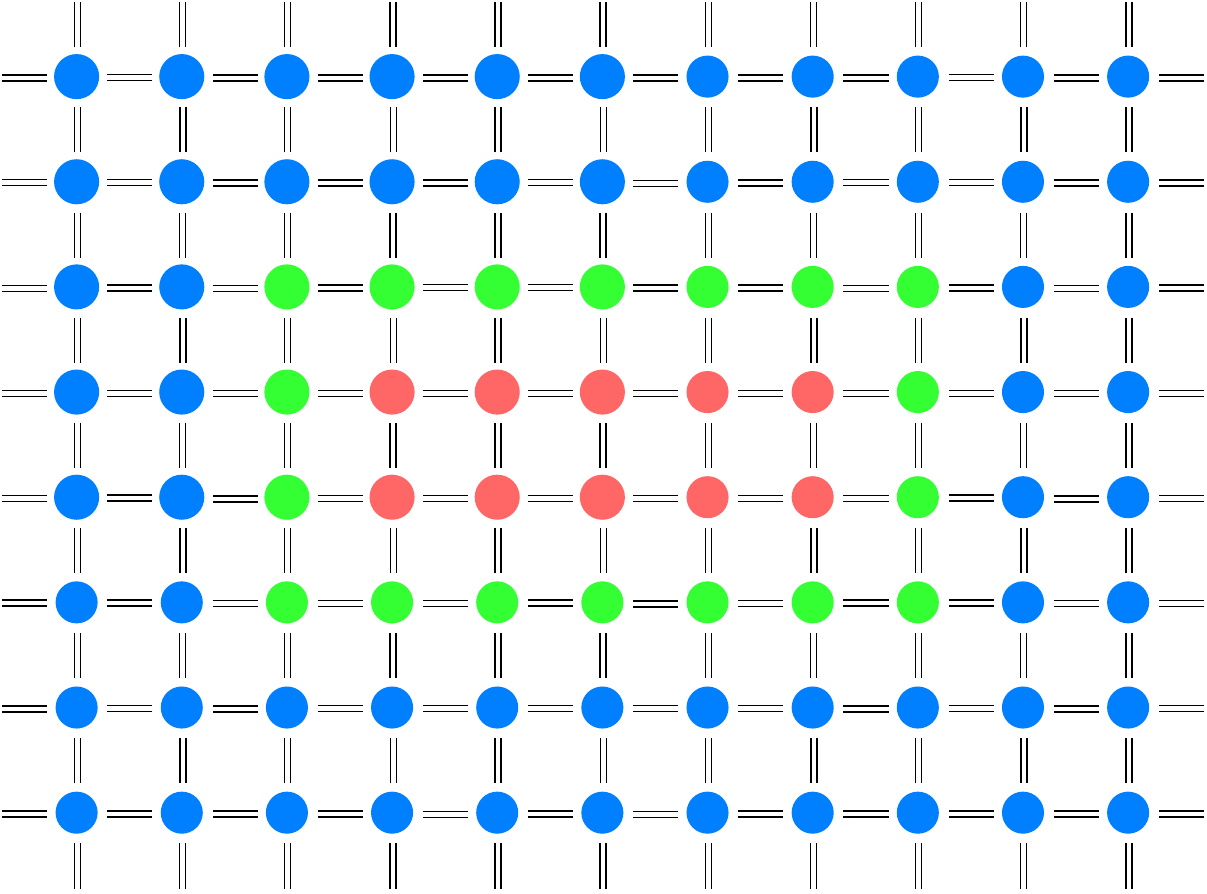}
    \caption{Example of the regions $\Lambda_1$, $\Lambda^\circ_1$, $\Lambda_2$, $\overline{\Lambda_2}$ and $\partial \Lambda$ on a two-dimensional lattice. The region $\Lambda_1$ consists of the union of all red and green lattice sites, while its interior $\Lambda^\circ_1$ is made up of all red sites only, cf. \eqref{eq:interior_lattice}. The exterior region $\Lambda_2$ consists of all blue lattice sites, while its closure $\overline{\Lambda_2}$ also includes the green sites. The boundary $\partial \Lambda$ is the set of all green lattice sites.}
    \label{fig:2D_lattice}
\end{figure}

Following \cite[Sec.~1.3.1]{Salmhofer2007}, we denote by $\varepsilon > 0$ the lattice spacing, by $L \gg \varepsilon$ the size of the lattice and require $L / \varepsilon \in \N$. Then, we define $\mathscr{L}^d_{\varepsilon,L} = \varepsilon \Z^d / L \Z^d$ to be our (finite) spatial $d$-dimensional lattice with periodic boundary conditions. Furthermore, for some region $\Omega \subsetneq \R^d / L \R^d$, let $\Lambda_1 = \Omega \cap \mathscr{L}^d_{\varepsilon, L}$ be the corresponding sublattice. Following \cite[Ch.~IV]{Guerra1975b}, we define a norm on $\Z^d$ via $\|n\|_\Z = \sum_{i=1}^d |n_i|$ and define the interior of the lattice $\Lambda_1$ as
\begin{equation}\label{eq:interior_lattice}
    \Lambda_1^\circ = \left\lbrace x = \varepsilon n \in \Lambda_1 \, : \, m \varepsilon \in \Lambda_1 \;\; \mathrm{if} \;\, \| n - m \|_\Z = 1 \right\rbrace \; .
\end{equation}
Finally, we define $\partial \Lambda = \Lambda_1 \setminus \Lambda^\circ_1$, $\Lambda_2 = \mathscr{L}^d_{\varepsilon, L} \setminus \Lambda_1$ and $\overline{\Lambda_2} = \Lambda_2 \cup \partial \Lambda = \mathscr{L}^d_{\varepsilon, L} \setminus \Lambda_1^\circ$ to be the boundary of $\Lambda_1$, the complement (or exterior to) $\Lambda_1$, and the ``closure'' of the complement of $\Lambda_1$, respectively, cf. Fig. \ref{fig:2D_lattice}.

Denote by $\ell_\R^2 (\mathscr{L}^d_{\varepsilon,L}, \varepsilon^d)$ the real Hilbert space of real-valued functions on the lattice $\mathscr{L}^d_{\varepsilon,L}$ with inner product $\braket{.,.}_\varepsilon$ given by
\begin{equation}
    \braket{\psi,\chi}_\varepsilon = \int_\mathscr{L} \D^d x \; \psi_x \chi_x \; ,
\end{equation}
where, following \cite[Sec.~1.3.1]{Salmhofer2007}, we introduced the suggestive notation
\begin{equation}
    \int_\mathscr{L} \D^d x \coloneqq \varepsilon^d \sum_{x \in \mathscr{L}^d_{\varepsilon,L}} \; .
\end{equation}
We define the quadratic form $\mathfrak{q}_\varepsilon$ on $\ell^2_\R (\mathscr{L}^d_{\varepsilon,L}, \varepsilon^d)$ as
\begin{equation}
    \mathfrak{q}_\varepsilon (\psi) \coloneqq \int_\mathscr{L} \D^d x \; ( \nabla_\varepsilon \psi_x )^2 \; ,
\end{equation}
where $\nabla_\varepsilon$ denotes the lattice gradient defined as \cite[Eq.~1.64]{Salmhofer2007}
\begin{equation}
    (\nabla_\varepsilon \psi)_{x,k} = \frac{1}{\varepsilon} \left( \psi_{x + \varepsilon \hat{e}_k} - \psi_x \right) \; , \quad\; k \in \{ 1, \ldots, d \} \; ,
\end{equation}
where $\hat{e}_k$ is the $k$-th unit vector in $\Z^d$.

The self-adjoint operator associated with the form $\mathfrak{q}_\varepsilon$ is the \emph{lattice Laplacian} $- \Delta_\varepsilon \coloneqq \nabla^*_\varepsilon \nabla_\varepsilon$, where the adjoint gradient $\nabla^*_\varepsilon$ is defined as the backwards difference operator acting as \cite[Eq.~1.71]{Salmhofer2007}
\begin{equation}
    (\nabla^*_\varepsilon \psi)_{x,k} = \frac{1}{\varepsilon} \left( \psi_{x - \varepsilon \hat{e}_k} - \psi_x \right) \; , \quad\; k \in \{ 1, \ldots, d \} \; .
\end{equation}
The lattice Laplacian acts on functions in $\ell_\R^2 (\mathscr{L}^d_{\varepsilon,L}, \varepsilon^d)$ as
\begin{equation}
    \left( - \Delta_\varepsilon \psi \right)_x = \frac{1}{\varepsilon^2} \sum_{k=1}^d \left( 2 \psi_x - \psi_{x + \varepsilon \hat{e}_k} - \psi_{x - \varepsilon \hat{e}_k} \right) \; .
\end{equation}
Moreover, $- \Delta_\varepsilon$ can be represented by a $|\mathscr{L}^d_{\varepsilon, L}| \times |\mathscr{L}^d_{\varepsilon, L}|$ matrix $- \hat{\Delta}_\varepsilon$, interpreted as a discrete integral kernel, with entries
\begin{equation}
    (- \hat{\Delta}_\varepsilon)_{xy} = \frac{1}{\varepsilon^2} \sum_{k=1}^d \left( 2 \delta^\varepsilon_{xy} - \delta^\varepsilon_{x + \varepsilon \hat{e}_k,y} - \delta^\varepsilon_{x - \varepsilon \hat{e}_k,y} \right) \; ,
\end{equation}
where $\delta^\varepsilon_{xy} \coloneqq \varepsilon^{-d} \delta_{xy}$ is the lattice Dirac-$\delta$. In particular, we have
\begin{equation}
    (- \Delta_\varepsilon \psi)_x = \int_\mathscr{L} \D^d x \; (- \hat{\Delta}_\varepsilon)_{xy} \, \psi_y \; .
\end{equation}
In order to illustrate the difference between $- \Delta_\varepsilon$ and $- \hat{\Delta}_\varepsilon$, we consider the limit $\varepsilon \to 0$, which formally yields
\begin{align}
    - \Delta_\varepsilon \xrightarrow{\varepsilon \to 0} - \Delta \; , \quad - \hat{\Delta}_\varepsilon \xrightarrow{\varepsilon \to 0} - \Delta \, \delta^{(d)} (x-y) \; ,
\end{align}
where $- \Delta$ is the continuum Laplacian and $\delta^{(d)} (x-y)$ is the $d$-dimensional Dirac-$\delta$ distribution. Therefore, we can indeed interpret $- \hat{\Delta}_\varepsilon$ as a discrete integral kernel with respect to the inner product of $\ell_\R^2 (\mathscr{L}^d_{\varepsilon,L}, \varepsilon^d)$, while $- \Delta_\varepsilon$ is the corresponding operator acting on functions.

Let $\Lambda_1 \subsetneq \mathscr{L}^d_{\varepsilon, L}$ be a sublattice with boundary $\partial \Lambda$ and complement $\Lambda_2$. For every $\psi \in \ell_\R^2 (\mathscr{L}^d_{\varepsilon, L}, \varepsilon^d)$, we define $\psi_1$, $\psi_2$ and $\underline{\psi_2}$ to be the restrictions of $\psi$ to $\Lambda_1$, $\Lambda_2$ and $\overline{\Lambda_2}$, respectively. We now wish to find a decomposition of the quadratic form $\mathfrak{q}_\varepsilon$ in three parts, one for the interior, one for the exterior and one coupling the interior and the exterior via their common boundary.

\begin{figure}[t]
    \centering
    \includegraphics[width=0.35\textwidth]{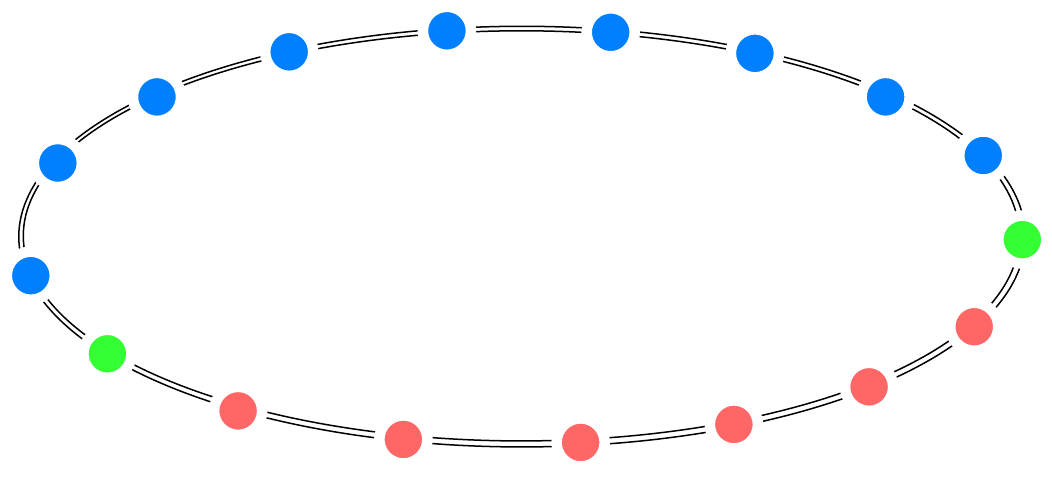}
    \caption{Sketch of a one-dimensional lattice with periodic boundary conditions. The color-coding is as in Fig. \ref{fig:2D_lattice}. In particular, the interior region $\Lambda_1$ (union of green and red lattice sites) corresponds to a single interval $\Omega \subsetneq \R / L \R$ with $N_1$ lattice sites, while the exterior region $\Lambda_2$ (blue lattice sites) corresponds to the complement of $\Lambda_1$ with $N_2$ lattice sites. The boundary $\partial \Lambda$ consists of the green lattice sites, and we denote the endpoints of the interval by $a$ and $b$, respectively, i.e., $\partial \Lambda \coloneqq \{a,b\}$.}
    \label{fig:chain_of_oscillators}
\end{figure}

It is instructive to start with a one-dimensional model. Let $\mathscr{L}^1_{\varepsilon,L}$ be a one-dimensional periodic lattice with $N \in \N$ lattice sites. Furthermore, let $\Lambda_1$ be the sublattice corresponding to an interval $\Omega \subsetneq \R / L \R$ with $N_1$ lattice sites and $\Lambda_2$ be the complement of $\Lambda_1$ with $N_2$ lattice sites, cf. Fig. \ref{fig:chain_of_oscillators}. Then, the kernel of the lattice Laplacian $- \hat{\Delta}_\varepsilon$ on $\ell^2_\R (\mathscr{L}^1_{\varepsilon,L}, \varepsilon)$ can be written as an $N \times N$ matrix given by
\begin{equation}
    - \hat{\Delta}_\varepsilon = \frac{1}{\varepsilon} \frac{1}{\varepsilon^2} \begin{pmatrix}
        2 & -1 & 0 & \cdots & 0 & 0 & -1 \\
        -1 & 2 & -1 & \cdots & 0 & 0 & 0 \\
        0 & -1 & 2 & \cdots & 0 & 0 & 0 \\
        \vdots & \vdots & \vdots & \ddots & \vdots & \vdots & \vdots \\
        0 & 0 & 0 & \cdots & 2 & -1 & 0 \\
        0 & 0 & 0 & \cdots & -1 & 2 & -1 \\
        -1 & 0 & 0 & \cdots & 0 & -1 & 2
    \end{pmatrix} \; .
\end{equation}
The corresponding quadratic form $\mathfrak{q}_\varepsilon$ is given by
\begin{equation}
    \mathfrak{q}_\varepsilon (\psi) = \iint_\mathscr{L} \D x \, \D y \; \psi_x (- \hat{\Delta}_\varepsilon)_{xy} \psi_y \; .
\end{equation}

Next, we define the $N_1 \times N_1$ matrix $- \hat{\Delta}^1_\varepsilon$,
\begin{equation}
    - \hat{\Delta}^1_\varepsilon = \frac{1}{\varepsilon} \frac{1}{\varepsilon^2} \begin{pmatrix}
        2 & -1 & 0 & \cdots & 0 & 0 & 0 \\
        -1 & 2 & -1 & \cdots & 0 & 0 & 0 \\
        0 & -1 & 2 & \cdots & 0 & 0 & 0 \\
        \vdots & \vdots & \vdots & \ddots & \vdots & \vdots & \vdots \\
        0 & 0 & 0 & \cdots & 2 & -1 & 0 \\
        0 & 0 & 0 & \cdots & -1 & 2 & -1 \\
        0 & 0 & 0 & \cdots & 0 & -1 & 2
    \end{pmatrix} \; ,
\end{equation}
the $N_2 \times N_2$ matrix $- \hat{\Delta}^2_\varepsilon$,
\begin{equation}
    - \hat{\Delta}^2_\varepsilon = \frac{1}{\varepsilon} \frac{1}{\varepsilon^2} \begin{pmatrix}
        2 & -1 & 0 & \cdots & 0 & 0 & 0 \\
        -1 & 2 & -1 & \cdots & 0 & 0 & 0 \\
        0 & -1 & 2 & \cdots & 0 & 0 & 0 \\
        \vdots & \vdots & \vdots & \ddots & \vdots & \vdots & \vdots \\
        0 & 0 & 0 & \cdots & 2 & -1 & 0 \\
        0 & 0 & 0 & \cdots & -1 & 2 & -1 \\
        0 & 0 & 0 & \cdots & 0 & -1 & 2
    \end{pmatrix} \; ,
\end{equation}
which differs from $- \hat{\Delta}^1_\varepsilon$ only in its dimension, and the $(N_2 + 2) \times (N_2 + 2)$ matrix $- \hat{\Delta}^{2, \star}_\varepsilon$,
\begin{equation}
    - \hat{\Delta}^{2, \star}_\varepsilon = \frac{1}{\varepsilon} \frac{1}{\varepsilon^2} \begin{pmatrix}
        0 & -1 & 0 & \cdots & 0 & 0 & 0 \\
        -1 & 2 & -1 & \cdots & 0 & 0 & 0 \\
        0 & -1 & 2 & \cdots & 0 & 0 & 0 \\
        \vdots & \vdots & \vdots & \ddots & \vdots & \vdots & \vdots \\
        0 & 0 & 0 & \cdots & 2 & -1 & 0 \\
        0 & 0 & 0 & \cdots & -1 & 2 & -1 \\
        0 & 0 & 0 & \cdots & 0 & -1 & 0
    \end{pmatrix} \; ,
\end{equation}
where the first and last diagonal entry are set to zero. By a straightforward calculation, we find that the quadratic form $\mathfrak{q}_\varepsilon$ can be written as
\begin{equation}
    \mathfrak{q}_\varepsilon (\psi) = \mathfrak{q}^1_\varepsilon (\psi_1) + \mathfrak{q}^{2, \star}_\varepsilon (\underline{\psi_2}) \; ,
\end{equation}
where $\mathfrak{q}^1_\varepsilon$ and $\mathfrak{q}^{2,\star}_\varepsilon$ are the quadratic forms associated with the lattice Laplacians $- \Delta^1_\varepsilon$ and $- \Delta^{2,\star}_\varepsilon$, respectively, i.e.,
\begin{align}
    \mathfrak{q}^1_\varepsilon (\psi_1) &= \iint_{\Lambda_1} \D x \, \D y \; \psi_{1,x} \, (- \hat{\Delta}^1_\varepsilon)_{xy} \, \psi_{1,y} \; , \\
    \mathfrak{q}^{2,\star}_\varepsilon (\underline{\psi_2}) &= \iint_{\overline{\Lambda_2}} \D x \, \D y \; \underline{\psi_{2,x}} \, (- \hat{\Delta}^{2,\star}_\varepsilon)_{xy} \, \underline{\psi_{2,y}} \; .
\end{align}
Let $\partial \Lambda \coloneqq \{ a,b \}$ be the boundary of the lattice $\Lambda_1$ and define $\partial \Lambda' \coloneqq \{ a + \varepsilon \hat{n}, b + \varepsilon \hat{n} \}$, where $\hat{n}$ is the unit outward (with respect to $\Lambda_1$) normal vector to the boundary $\partial \Lambda$. Notice that the Laplacian $- \Delta^1_\varepsilon$ describes \emph{homogeneous} Dirichlet boundary conditions on $\partial \Lambda'$, while the Laplacian $- \Delta^{2,\star}_\varepsilon$ describes \emph{inhomogeneous} Dirichlet boundary conditions given by the value of $f_1$ on the boundary $\partial \Lambda$. More precisely, following, e.g., \cite[Sec.~2.3.1]{Pulliam2014}, we introduce the $N_2$-component vector $b_\psi$ as
\begin{equation}\label{eq:boundary_value_vector}
    b_\psi^\mathsf{T} \coloneqq \frac{1}{\varepsilon^2} \left( \psi_{a}, 0, 0, \ldots, 0, 0, \psi_{b} \right) \; .
\end{equation}
Then, by direct calculation, it can be shown that
\begin{equation}
    \left. - \Delta^{2,\star}_\varepsilon \underline{\psi_2} \right|_{\Lambda_2} = - \Delta^2_\varepsilon \psi_2 - b_\psi \; .
\end{equation}
Similarly, the quadratic form $\mathfrak{q}^{2,\star}_\varepsilon$ can be written as
\begin{equation}
    \mathfrak{q}^{2,\star}_\varepsilon (\underline{\psi_2}) = \mathfrak{q}^2_\varepsilon (\psi_2) - 2 \braket{\psi_2, b_\psi}_\varepsilon \; ,
\end{equation}
where $\mathfrak{q}^2_\varepsilon$ is the quadratic form associated with the Laplacian $- \Delta^2_\varepsilon$, i.e., the quadratic form in $\Lambda_2$ with homogeneous Dirichlet boundary conditions on $\partial \Lambda$.

In summary, we have found that the quadratic form $\mathfrak{q}_\varepsilon$ can be split as
\begin{equation}\label{eq:quadratic_form_splitting}
    \mathfrak{q}_\varepsilon (\psi) = \mathfrak{q}^1_\varepsilon (\psi_1) + \mathfrak{q}^2_\varepsilon (\psi_2) - 2 \braket{\psi_2, b_\psi}_\varepsilon \; ,
\end{equation}
where $\mathfrak{q}^1_\varepsilon$ and $\mathfrak{q}^2_\varepsilon$ are the quadratic forms associated with the Laplacians $- \Delta^1_\varepsilon$ and $- \Delta^2_\varepsilon$, respectively, and $b_\psi$ is the vector of boundary values of $\psi_1$. Notice that $\mathfrak{q}^1_\varepsilon$ and $\mathfrak{q}^2_\varepsilon$ only contain degrees of freedom in $\Lambda_1$ and $\Lambda_2$, respectively, while the last term linearly couples the two regions across boundary $\partial \Lambda$. This result generalizes to higher dimensions in a straightforward way.

\subsection{Example: Euclidean Field Theory}\label{sec:Euclidean_Field_Theory}

In order to further elucidate the above splitting of the quadratic form $\mathfrak{q}_\varepsilon$, we now consider a Euclidean field theory on the lattice. The purpose of this Section is to recover a Euclidean field theory on a subregion with \emph{free} boundary conditions as discussed in \cite{Guerra1975a,Guerra1975b,Guerra1976,Floerchinger2023}, thereby testing the consistency of the splitting discussed in Section \ref{sec:Lattice_Laplacian}.

Let $\mathscr{L}^1_{\varepsilon, L}$ be a one-dimensional periodic lattice and $\Lambda_1$ and $\Lambda_2$ be two sublattices corresponding to intervals as defined in Section \ref{sec:Lattice_Laplacian}. Then, the classical lattice field theory is described by the quadratic form $S_\mathrm{E}$, called the Euclidean action, given by
\begin{equation}\label{eq:Euclidean_action}
    S_\mathrm{E} (\psi) = \frac{1}{2} \iint_\mathscr{L} \D x \, \D y \; \psi_x \, (- \hat{\Delta}_\varepsilon + m^2 \delta^\varepsilon )_{xy} \, \psi_y \; ,
\end{equation}
where $m > 0$ is the mass parameter of the theory.

Since the mass term $m^2 \delta^\varepsilon$ does not couple neighbouring lattice sites, we can, analogues to the case of the form $\mathfrak{q}_\varepsilon$, split the Euclidean action $S_\mathrm{E}$ into three parts, one for the interior, one for the exterior and one coupling the interior and the exterior via a common boundary. More precisely, we have\footnote{Note that a factor of $2$ is missing compared to \eqref{eq:quadratic_form_splitting} due to the factor of $1/2$ in the definition of the Euclidean action in \eqref{eq:Euclidean_action}.}
\begin{equation}\label{eq:Euclidean_action_splitting}
    S_\mathrm{E} (\psi) = S^1_\mathrm{E} (\psi_1) + S^2_\mathrm{E} (\psi_2) - \braket{\psi_2, b_\psi}_\varepsilon \; ,
\end{equation}
where $S^1_\mathrm{E}$ and $S^2_\mathrm{E}$ are the Euclidean actions associated with the Laplacians $- \Delta^1_\varepsilon$ and $- \Delta^2_\varepsilon$, respectively, and $b_\psi$ is the vector of boundary values of $\psi_1$ as introduced in \eqref{eq:boundary_value_vector}. Notice that from the point of view of the field in region $\Lambda_2$, $- \braket{\psi_2, b_\psi}_\varepsilon$ is a source term and the boundary values of the field in $\Lambda_1$ act as an external source supported exclusively on the boundary.

The Euclidean action $S_\mathrm{E}$ defines a classical statistical lattice field theory via a centred Gaussian measure $\mu$ on $\R^N$ given by
\begin{equation}
    \D \mu (\psi) \coloneqq \frac{1}{Z} \E^{- S_\mathrm{E} (\psi)} \, \D^N \psi \; ,
\end{equation}
where $Z$ is a normalization constant and $\D^N \psi$ is the $N$-dimensional Lebesgue measure. The measure $\mu$ is completely characterized by its covariance operator $\hat{G}$ given by
\begin{align}
    \hat{G} &= \varepsilon^{-2} (- \hat{\Delta}_\varepsilon + m^2 \delta^\varepsilon)^{-1} \; , \\
    G_{xy} &\coloneqq (\hat{G})_{xy} = \int_{\R^N} \psi_x \psi_y \; \D \mu (\psi) \; .
\end{align}

Upon splitting the action into three parts, the probability density with respect to $\D^N \psi = \D^{N_1} \psi_1 \, \D^{N_2} \psi_2$ factorizes as
\begin{equation}
    \begin{split}
        &\D \mu (\psi) = \D \mu (\psi_1, \psi_2) \\
        &= \frac{1}{Z} \, \E^{- \left( S^1_\mathrm{E} (\psi_1) + S^2_\mathrm{E} (\psi_2) - \braket{\psi_2, b_\psi}_\varepsilon \right)} \; \D^{N_1} \psi_1 \, \D^{N_2} \psi_2 ,
    \end{split}
\end{equation}
We obtain a reduced theory for the region $\Lambda_1$ by considering the marginal $\mu_1$ given by integrating out the degrees of freedom in $\Lambda_2$, i.e.,
\begin{equation}
    \begin{split}
        &\D \mu_1 (\psi_1) = \int_{\R^{N_2}} \D \mu (\psi_1, \psi_2) \\
        &= \frac{1}{Z} \left( \int_{\R^{N_2}} \E^{- \left( S^2_\mathrm{E} (\psi_2) - \braket{\psi_2, b_\psi}_\varepsilon \right)} \, \D^{N_2} \psi_2 \right) \E^{- S^1_\mathrm{E} (\psi_1)} \; \D^{N_1} \psi_1 \\
        &= \frac{1}{Z} \sqrt{\det \left( 2 \pi \hat{G}_{\mathrm{D},2} \right)} \; \E^{- \widetilde{S^1_\mathrm{E}} (\psi_1)} \; \D^{N_1} \psi_1 \; ,
    \end{split}
\end{equation}
where $\hat{G}_{\mathrm{D},2}$ is the covariance operator\footnote{The index $\mathrm{D}$ indicates the Dirichlet boundary conditions described by the Euclidean action $S^2_\mathrm{E}$.} associated with the Euclidean action $S^2_\mathrm{E}$, i.e.,
\begin{equation}
    \hat{G}_{\mathrm{D},2} \coloneqq \varepsilon^{-2} \, (- \hat{\Delta}^2_\varepsilon + m^2 \delta^\varepsilon)^{-1} \; ,
\end{equation}
and $\widetilde{S^1_\mathrm{E}}$ is a quadratic form given by
\begin{equation}
    \begin{split}
        &\widetilde{S^1_\mathrm{E}} (\psi_1) = S^1_\mathrm{E} (\psi_1) - \frac{\varepsilon^2}{2} b_\psi^\mathsf{T} \, \hat{G}_{\mathrm{D,2}} \, b_\psi \\
        &= S^1_\mathrm{E} (\psi_1) - \frac{1}{2} \sum_{x,y \in \partial \Lambda} \psi_{1,x} \, \frac{(\hat{G}_{\mathrm{D},2})_{x + \varepsilon \hat{n}, y + \varepsilon \hat{n}}}{\varepsilon^2} \, \psi_{1,y} \; ,
    \end{split}
\end{equation}
where $\hat{n}$ is again the unit outward normal vector with respect to $\Lambda_1$. Since $\hat{G}_{\mathrm{D},2}$ is a discrete Dirichlet Green's function, it is natural to extend it to a $(N_2+2) \times (N_2+2)$ matrix, which we again denote by $\hat{G}_{\mathrm{D},2}$, whose entries are zero if $x \in \partial \Lambda \vee y \in \partial \Lambda$. Thus, we see that
\begin{equation}
    \frac{(\hat{G}_{\mathrm{D},2})_{x + \varepsilon \hat{n}, y + \varepsilon \hat{n}}}{\varepsilon^2} = (\partial_{n_x}^\varepsilon \partial_{n_y}^\varepsilon \hat{G}_{\mathrm{D},2})_{xy}
\end{equation}
for $x,y \in \partial \Lambda$, where we introduced the discrete (outward with respect to $\Lambda_1$) normal derivative $\partial_n^\varepsilon \coloneqq \hat{n} \cdot \nabla_\varepsilon$.

Thus, we finally obtain
\begin{equation}\label{eq:tilde_euclidean_action}
    \begin{split}
        \widetilde{S^1_\mathrm{E}} (\psi_1) &= S^1_\mathrm{E} (\psi_1) - \frac{1}{2} \sum_{x,y \in \partial \Lambda} \psi_{1,x} \, (\partial_{n_x}^\varepsilon \partial_{n_y}^\varepsilon \hat{G}_{\mathrm{D},2})_{xy} \, \psi_{1,y} \\
        &= \frac{1}{2} \iint_{\Lambda_1} \D x \, \D y \; \psi_{1,x} \, (- \hat{\Delta}_\varepsilon^{1,m} + m^2 \delta^\varepsilon )_{xy} \, \psi_{1,y} \; ,
    \end{split}
\end{equation}
where $- \Delta^{1,m}_\varepsilon$ is the Laplacian on $\Lambda_1$ with non-local and mass dependent boundary conditions (see discussion in the remainder of this Section) such that the above equality holds \cite{Guerra1976,Floerchinger2023}. More precisely, $- \hat{\Delta}^{1,m}_\varepsilon$ is defined as
\begin{equation}
    - \hat{\Delta}^{1,m}_\varepsilon \coloneqq - \hat{\Delta}^1_\varepsilon - B^m_{\partial \Lambda} \; ,
\end{equation}
where $B^m_{\partial \Lambda}$ is a (mass dependent) matrix ``concentrated on the boundary'' \cite[Sec.~IV.2]{Guerra1975b}, see also \cite[Thm.~IV.7]{Guerra1975b}. In particular, we have
\begin{equation}
    (B^m_{\partial \Lambda})_{xy} = \frac{1}{\varepsilon^2} \begin{cases}
        (\partial_{n_x}^\varepsilon \partial_{n_y}^\varepsilon \hat{G}_{\mathrm{D},2})_{xy} \quad & \mathrm{if} \; x,y \in \partial \Lambda \; , \\
        0 & \mathrm{otherwise} \; .
    \end{cases}
\end{equation}
The mass dependence of the boundary term is a consequence of the mass dependence of the Dirichlet Green's matrix $\hat{G}_{\mathrm{D},2}$. In $d$ dimensions, the above expression generalizes to\footnote{We assume that the region $\Lambda_1$ is chosen such that the normal unit vector $\hat{n}$ is well-defined for all boundary points.}
\begin{equation}
    \begin{split}
        \widetilde{S^1_\mathrm{E}} (\psi_1) &= \frac{1}{2} \iint_{\Lambda_1} \D^d x \, \D^d y \; \psi_{1,x} \, (- \hat{\Delta}^1_\varepsilon + m^2 \delta^\varepsilon )_{xy} \, \psi_{1,y} \\
        &- \frac{1}{2} \iint_{\partial \Lambda} \D S (x) \, \D S (y) \; \psi_{1,x} \, (\partial_{n_x}^\varepsilon \partial_{n_y}^\varepsilon \hat{G}_{\mathrm{D},2})_{xy} \, \psi_{1,y} \; ,
    \end{split}
\end{equation}
where we introduced the notation
\begin{equation}
    \int_{\partial \Lambda} \D S (x) \coloneqq \varepsilon^{d-1} \sum_{x \in \partial \Lambda} \; .
\end{equation}

We now further elaborate on the boundary conditions described by the Laplacian $- \Delta^{1,m}_\varepsilon$. In order to do so, we derive the (lattice regularized) partial differential equation associated to the Euclidean action $\widetilde{S^1_\mathrm{E}}$ via the principle of least action. Varying $\widetilde{S^1_\mathrm{E}}$ with respect to $\psi_1$ and setting the result to zero, we obtain the equation
\begin{equation}
    \int_{\Lambda_1} \D y \; \left( - \hat{\Delta}^1_\varepsilon + m^2 \delta^\varepsilon - B^m_{\partial \Lambda} \right)_{xy} \psi_{1,y} \overset{!}{=} 0 \; .
\end{equation}
Since this equation must hold for all $\psi_1 \in \ell_\R^2 (\Lambda_1, \varepsilon^d)$, we find the set of equations
\begin{equation}\label{eq:laplace_equation_euclidean_FT}
    \left( \left( - \Delta_\varepsilon + m^2 \right) \psi_1 \right)_x = 0 \; ,
\end{equation}
for $x \in \Lambda^\circ_1$ and
\begin{equation}\label{eq:boundary_condition_euclidean_FT}
    \left( \partial^\varepsilon_n \psi_1 \right)_x + \varepsilon m^2 \psi_{1,x} = \int_{\partial \Lambda} \D S (y) \; \widetilde{k}_{xy} \, \psi_{1,y} - \frac{\psi_{1,x}}{\varepsilon} \; ,
\end{equation}
for $x \in \partial \Lambda$, where $\widetilde{k} \coloneqq \partial_{n_x}^\varepsilon \partial_{n_y}^\varepsilon \hat{G}_{\mathrm{D},2}$.

Notice that the term $- \psi_{1,x} / \varepsilon$ on the right-hand side of \eqref{eq:boundary_condition_euclidean_FT} stems from the fact that the lattice Laplacian in $S^1_\mathrm{E}$ describes Dirichlet boundary conditions on $\partial \Lambda'$. It is \emph{a priori} not clear how to interpret this term in the continuum limit. However, we observe that the term $- \psi_{1,x} / \varepsilon$ is necessary in order to ensure that the boundary conditions are well-defined in the continuum limit. For simplicity, we work in $d=1$ in the following. Then, the term $- \psi_{1,x} / \varepsilon$ is needed in order to ensure finite values of the right-hand side of \eqref{eq:boundary_condition_euclidean_FT} in the continuum limit. More precisely, the continuum limit of $\widetilde{k}_{xx} - \varepsilon^{-1}$ for $x \in \partial \Lambda$ yields the finite value\footnote{In the derivation of \eqref{eq:limit_double_normal_derivative} we used the expression for the Green's function of $- \Delta + m^2$ with Dirichlet boundary conditions on an interval as given in \cite[\S II.5]{Guerra1975a}.}
\begin{equation}\label{eq:limit_double_normal_derivative}
    \pm \lim_{x' \to x} \frac{\partial^2 G_{\mathrm{D},2}}{\partial x' \partial n_x} (x',x) = - m \coth (m L) \; ,
\end{equation}
$x \in \partial \Lambda$, where the sign on the left-hand side depends on the orientation of the normal vector $\hat{n}$ and $L$ is the length of the interval $\Omega_2$ corresponding to $\Lambda_2$.

We can also understand the need to subtract a divergent term by looking at the expression of the double normal derivative of $G_{\mathrm{D},2}$ as a distributional kernel. More precisely, in one dimension $\partial_{x} \partial_{y} G_{\mathrm{D},2}$ takes the form
\begin{equation}
    \begin{split}
        &\frac{\partial^2 G_{\mathrm{D},2}}{\partial x \partial y} (x,y) = \frac{1}{2} \E^{-m (x-y + |x-y|)} \bigg[ - m \E^{m |x-y|} \\
        &\times \left( \E^{2 m (x-y)} (\coth (m L) - 1) + \coth (m L) \right) \\
        &+ \E^{m |x-y|} \left( 2 \delta (x-y) - m \right) \bigg] \; ,
    \end{split}
\end{equation}
where we used $|x|' = \mathrm{sgn} (x)$ and $\mathrm{sgn} (x)' = 2 \delta (x)$ in the sense of distributions. Thus, we formally have
\begin{equation}
    \left. \frac{\partial^2 G_{\mathrm{D},2}}{\partial n_x \partial n_y} (x,y) \right|_{x=y \in \{0,L\}} = - m \coth (m L) + \delta (0) \; .
\end{equation}
Therefore, subtracting the term $\varepsilon^{-1}$ from $\partial_{n_x}^\varepsilon \partial_{n_y}^\varepsilon \hat{G}_{\mathrm{D},2}$ corresponds to subtracting the term $\delta (0)$ in the continuum limit, which ensures that the kernel on the right-hand side of \eqref{eq:boundary_condition_euclidean_FT} assumes the finite value \eqref{eq:limit_double_normal_derivative}, defined via a limit from within the interval of length $L$, in the continuum limit.

In summary, we see that in $d$ Euclidean spacetime dimensions, the continuum limit of the system of equations \eqref{eq:laplace_equation_euclidean_FT} and \eqref{eq:boundary_condition_euclidean_FT} yields the boundary value problem
\begin{subequations}
    \begin{alignat}{2}
        &(- \Delta + m^2) \psi_1 (x) = 0 \; , \qquad &&x \in \Omega \; , \\
        &\frac{\partial \psi_1}{\partial n} (x) = \int_{\partial \Omega} \D S (y) \; k (x,y) \, \psi_1 (y) \; , \quad\;\; &&x \in \partial \Omega \; , \label{eq:boundary_condition_euclidean_FT_continuum}
    \end{alignat}
\end{subequations}
where $k (x,y) \coloneqq (\partial^2 G_{\mathrm{D},2} / \partial n_x \partial n_y) (x,y)$ is defined as the limit from within the region $\Omega$. We observe that the boundary conditions in \eqref{eq:boundary_condition_euclidean_FT_continuum} are precisely the free boundary conditions as discussed in \cite{Guerra1975a,Guerra1975b,Guerra1976}, see also \cite{Floerchinger2023}. We conclude that the splitting of the Euclidean action given in \eqref{eq:Euclidean_action_splitting} yields the correct boundary conditions for the reduced theory in the interior region $\Lambda_1$.

In conclusion, the effect of integrating out the field on the exterior lattice $\Lambda_2$ is to induce non-local and mass dependent boundary conditions on the interior lattice $\Lambda_1$ such that the covariance matrix associated with the reduced theory is given by the restriction of the global covariance matrix to the interior lattice $\Lambda_1$. Of course, this is just a manifestation of the fact that the reduced theory describes, by construction, the same physics in the region $\Lambda_1$ as the full theory on the entire lattice $\mathscr{L}^d_{\varepsilon,L}$. That the effect of the exterior degrees of freedom is completely described by a boundary term is a consequence of the Markov property of the free scalar field, see the discussions in \cite{Nelson1973a,Nelson1973b,Guerra1975a,Guerra1975b,Guerra1976,Floerchinger2023}.

\subsection{Quantum Lattice System}\label{sec:Quantum_Lattice_System}

We now turn to quantum lattice systems representing lattice regularized scalar quantum field theories\footnote{For details on such systems, see, e.g., \cite{Nachtergaele2008,Bratteli1987,Bratteli1981,Simon1993}.}. Let $\mathscr{L}^d_{\varepsilon,L}$ be a $d$-dimensional periodic lattice as defined Section \ref{sec:Lattice_Laplacian}. We denote by $\hat{\Phi}_x$ and $\hat{\Pi}_x$ the (unbounded) field and conjugated momentum field operators, respectively, at the lattice site $x \in \mathscr{L}^d_{\varepsilon,L}$, satisfying the canonical commutation relations (CCRs)
\begin{align}
    \left[ \hat{\Phi}_x, \hat{\Phi}_y \right] &= \left[ \hat{\Pi}_x, \hat{\Pi}_y \right] = 0 \; , \\
    \left[ \hat{\Phi}_x, \hat{\Pi}_y \right] &= \I \delta^\varepsilon_{xy} \mathds{1} \; ,
\end{align}
for $x,y \in \mathscr{L}^d_{\varepsilon,L}$. The dynamics of the system are governed by the Hamiltonian of an anharmonic lattice defined as
\begin{equation}\label{eq:Hamiltonian_operator_anharmonic}
    \hat{H} \coloneqq \frac{1}{2} \int_{\mathscr{L}} \D^d x \left[ \hat{\Pi}_x^2 + \int_{\mathscr{L}} \D^d y \; \hat{\Phi}_x D^\varepsilon_{xy} \hat{\Phi}_y + V (\hat{\Phi}_x) \right] \, ,
\end{equation}
where we defined $D^\varepsilon \coloneqq - \hat{\Delta}_\varepsilon + m^2 \delta^\varepsilon$ and $V$ is some suitable potential bounded from below.

We define the Schr\"odinger representation of the lattice theory to be the representation of the CCR algebra on the representation Hilbert space $\mathcal{H} = \bigotimes_{x \in \mathscr{L}^d_{\varepsilon,L}} L^2 (\R, \D \varphi_x)$. In the Schr\"odinger representation, the Hamiltonian acts as the operator
\begin{equation}\label{eq:Hamiltonian_operator_Schroedinger}
    \hat{H} \coloneqq \frac{1}{2} \int_\mathscr{L} \D^d x \left[ - \frac{\partial^2}{\partial \varphi_x^2} + \int_\mathscr{L} \D^d y \; \varphi_x  D^\varepsilon_{xy} \varphi_y  + V (\varphi_x) \right] \, ,
\end{equation}
where $- \partial^2 / \partial \varphi_x^2$ denotes the unique self-adjoint realization of the Laplacian on $L^2 (\R, \mathrm{d} \varphi_x)$ \cite{Reed1981,Reed1975}.

Let $\Lambda_1 \subsetneq \mathscr{L}^d_{\varepsilon, L}$ be a sublattice with boundary $\partial \Lambda$ and complement $\Lambda_2$. Except for the term containing the lattice Laplacian, the Hamiltonian in \eqref{eq:Hamiltonian_operator_Schroedinger} is local in the lattice indices, and we may split it, following the discussion in Section \ref{sec:Lattice_Laplacian}, as $\hat{H} = \hat{H}_1 + \hat{H}_2 + \hat{H}_{12}$, where
\begin{align}
    \hat{H}_1 &= \frac{1}{2} \int_{\Lambda_1} \D^d x \left[ - \frac{\partial^2}{\partial \varphi_x^2} + \int_{\Lambda_1} \D^d y \; \varphi_x D^{\varepsilon,1}_{xy} \varphi_y + V (\varphi_x) \right] , \\
    \hat{H}_2 &= \frac{1}{2} \int_{\Lambda_2} \D^d x \left[ - \frac{\partial^2}{\partial \varphi_x^2} + \int_{\Lambda_2} \D^d y \; \varphi_x D^{\varepsilon,2}_{xy} \varphi_y + V (\varphi_x) \right] , \\
    \hat{H}_{12} &= - \int_{\partial \Lambda} \D S (x) \; \frac{\varphi_{x + \varepsilon \hat{n}} \, \varphi_x}{\varepsilon} \; , \label{eq:Hamiltonian_operator_split}
\end{align}
where $\hat{n}$ denotes again the outward normal vector (with respect to $\Lambda_1$) on the boundary $\partial \Lambda$ and $D^{\varepsilon,i} = - \hat{\Delta}_\varepsilon^i + m^2 \delta^\varepsilon$ for $i \in \{1,2\}$. Notice that the boundary potential $\hat{H}_{12}$ linearly couples the fields in $\Lambda_1$ and $\Lambda_2$ across $\partial \Lambda$.

\section{Local Dynamics of a Free Scalar Field}\label{sec:Local_Dynamics}

In this Section, we study the local dynamics of a relativistic quantum field theory using the concrete example of a free, massive scalar field. To do this, we use the Feynman-Vernon influence functional approach \cite{Feynman1963}, which is closely related to the Schwinger-Keldysh formalism \cite{Schwinger1961,Keldysh1964} (see also \cite{Berges2004,Berges2015}) and allows the study of the time evolution of open quantum systems out of equilibrium. A detailed introduction to these concepts can be found, e.g., in \cite{Calzetta2022}, which we will also follow in this paper.

Even though states in relativistic quantum field theories are generically entangled \cite{Witten2018,Hollands2018}, we will first consider a product ansatz for the initial state, i.e., a state that can be written as a tensor product of states in the two regions $\Lambda_1$ and $\Lambda_2$. This will allow us to elucidate the influence of the environment on the reduced dynamics of the system. After that, we incorporate initial state correlations by considering an initial state that is a (local) perturbation of a thermal state. Finally, we derive a stochastic equation of motion for the expectation value of the field in the interior region.

For definiteness, we consider a lattice model as introduced in Section \ref{sec:Lattice_Model}. We denote the field in $\Lambda_1$ by $\varphi$ and the field in $\Lambda_2$ by $\phi$, respectively. The dynamics of the theory are assumed to be governed by the Hamiltonian of a free massive scalar field on a lattice given by \eqref{eq:Hamiltonian_operator_anharmonic} with $V \equiv 0$. In particular, this Hamiltonian can be split as $\hat{H} = \hat{H}_1 + \hat{H}_2 + \hat{H}_{12}$, where $\hat{H}_1$ and $\hat{H}_2$ are the Hamiltonians associated with the regions $\Lambda_1$ and $\Lambda_2$, respectively, and $\hat{H}_{12}$ couples the two regions across the boundary $\partial \Lambda$, see \eqref{eq:Hamiltonian_operator_split}. In the language of open quantum systems, we may think of $\hat{H}_1$ as the system Hamiltonian, $\hat{H}_2$ as the environment Hamiltonian and $\hat{H}_{12}$ as the interaction Hamiltonian which couples the system to the environment.

The dynamics described by this Hamiltonian can also be described by the classical action $S$ given by
\begin{equation}
    S (\psi) = \frac{1}{2} \int_{0}^{t} \D \tau \int_{\mathscr{L}} \D^d x \left[ \dot{\psi}_x^2 - \int_{\mathscr{L}} \D^d y \; \psi_x \, D^\varepsilon_{xy} \, \psi_y \right] \, ,
\end{equation}
where $\dot{\psi}_x \coloneqq \partial \psi_x / \partial \tau$. In terms of the interior and exterior fields $\varphi$ and $\phi$, this action splits as
\begin{equation}
    S (\varphi, \phi) = S_1 (\varphi) + S_2 (\phi) + S_{12} (\varphi, \phi) \; ,
\end{equation}
where $S_1$ and $S_2$ are the action functionals associated with the regions $\Lambda_1$ and $\Lambda_2$, respectively, given by
\begin{equation}
    S_i (\psi) = \frac{1}{2} \int_{0}^{t} \D \tau \int_{\Lambda_i} \D^d x \left[ \dot{\psi}_x^2 - \int_{\Lambda_i} \D^d y \; \psi_x \, D^{\varepsilon,i}_{xy} \, \psi_y \right]
\end{equation}
for $i \in \{ 1,2 \}$ and $S_{12}$ couples the fields at the boundary, i.e.,
\begin{equation}\label{eq:S_12}
    S_{12} (\varphi, \phi) = - \int_{0}^{t} \D \tau \int_{\partial \Lambda} \D S (x) \; \frac{\phi_{x + \varepsilon \hat{n}} \, \varphi_x}{\varepsilon} \; .
\end{equation}

\subsection{Factorizing Initial State}\label{sec:Factorizing_Initial_State}

\begin{figure}[t]
    \centering
    \includegraphics[width=0.4\textwidth]{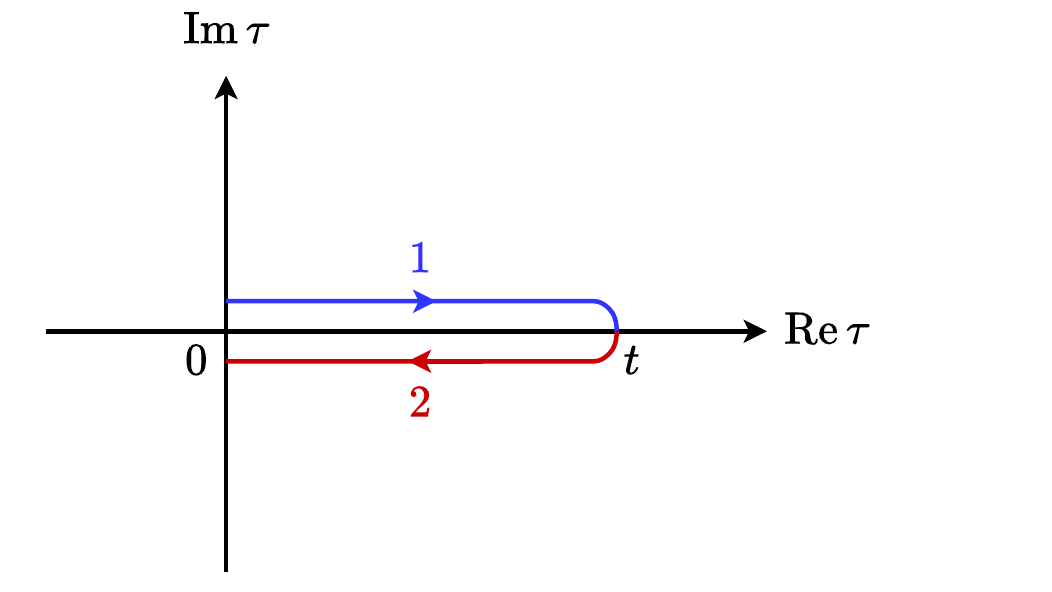}
    \caption[Schwinger-Keldysh contour $\mathscr{C}$ for the influence functional.]{Schwinger-Keldysh contour $\mathscr{C}$ for the influence functional. The contour starts at $\tau = 0$, extends to $\tau = t$ and ends again at $\tau = 0$. The contour is composed of two branches. Branch $1$ is the forward branch and branch $2$ is the backward branch as indicated by the arrows in the sketch. For illustration purposes, the two branches are deformed into the complex plane.}
    \label{fig:SK_contour}
\end{figure}

We start by assuming that the initial state of the system factorizes with respect to the splitting into regions $\Lambda_1$ and $\Lambda_2$. More precisely, we assume that the initial state of the system is described by a density operator of the form $\hat{\rho} (0) = \hat{\rho}_1 (0) \otimes \hat{\rho}_2 (0)$, where $\hat{\rho}_1 (0)$ and $\hat{\rho}_2 (0)$ are the initial states of the regions $\Lambda_1$ and $\Lambda_2$, respectively. In the Schr\"odinger representation, such a product state can be written in terms of density kernels as
\begin{equation}\label{eq:product_state}
    \rho (0; \varphi^1, \phi^1, \varphi^2, \phi^2) = \rho_1 (0; \varphi^1, \varphi^2) \, \rho_2 (0; \phi^1, \phi^2) \; .
\end{equation}
In this case, the influence functional \cite{Feynman1963}, describing the non-unitary contribution to the time evolution, is given by \cite[Sec.~3.2]{Calzetta2022}
\begin{equation}\label{eq:influence_functional_product_state}
    \begin{split}
        &\mathfrak{F}_\mathrm{IF} (\varphi^1, \varphi^2) = \E^{\I S_\mathrm{IF} (\varphi^1, \varphi^2, t)} \\
        &= \int_\mathscr{C} \mathcal{D} \phi \; \E^{\I (S_2 (\phi) + S_{12} (\varphi, \phi))} \rho_2 (0; \phi_0^1, \phi_0^2) \; ,
    \end{split}
\end{equation}
where $S_\mathrm{IF}$ is called the influence action and $\mathscr{C}$ is Schwinger-Keldysh contour shown in Fig. \ref{fig:SK_contour}.

Consider now, for simplicity, Gaussian initial conditions in the environment. In this case, the Gaussian integrals in \eqref{eq:influence_functional_product_state} can be performed exactly, and we obtain for the influence action \cite[Sec.~3.2.2]{Calzetta2022}
\begin{equation}\label{eq:influence_action_product_state}
    \begin{split}
        &S_\mathrm{IF} (\varphi^1, \varphi^2, t) = \\
        &\frac{\I}{2} \int_0^t \D \tau \, \D \tau' \int_{\partial \Lambda} \D S (x) \, \D S (y) \; \varphi^a_x (\tau) \, ( \mathcal{K}_{xy} )_{ab} (\tau, \tau') \, \varphi^b_y (\tau') \; ,
    \end{split}
\end{equation}
where summation over repeated ``time-path'' indices $a,b \in \{1,2\}$ is implied and $\mathcal{K}$ is the path ordered propagator given by
\begin{equation}\label{eq:path_ordered_propagator_product_state}
    \begin{split}
        &\mathcal{K}_{xy} (\tau, \tau') \coloneqq \\
        &\varepsilon^{-2} \begin{pmatrix}
            \braket{\mathcal{T} \phi_{x + \varepsilon \hat{n}} (\tau) \, \phi_{y + \varepsilon \hat{n}} (\tau')} & - \braket{\phi_{y + \varepsilon \hat{n}} (\tau') \, \phi_{x + \varepsilon \hat{n}} (\tau)} \\
            - \braket{\phi_{x + \varepsilon \hat{n}} (\tau) \, \phi_{y + \varepsilon \hat{n}} (\tau')} & \braket{\widetilde{\mathcal{T}} \phi_{x + \varepsilon \hat{n}} (\tau) \, \phi_{y + \varepsilon \hat{n}} (\tau')}
        \end{pmatrix} \; .
    \end{split}
\end{equation}
Here, $\mathcal{T}$ denotes time-ordering (the latest time to the left), $\widetilde{\mathcal{T}}$ denotes anti time-ordering (the latest time to the right) and the expectation values $\braket{\cdot}$ are meant with respect to the path integral of the region $\Lambda_2$ \emph{without} the coupling to the field in $\Lambda_1$. More precisely, the expectation values are given by
\begin{equation}
    \braket{\mathcal{O} (t')} = \int_\mathscr{C} \mathcal{D} \phi \; \E^{\I S_2 (\phi)} \, \mathcal{O} (t') \, \rho_2 (0; \phi^1_0, \phi^2_0) \; .
\end{equation}

We observe from \eqref{eq:influence_action_product_state} that the influence action is a boundary action. This is a consequence of the fact that the Laplacian linearly couples the interior and exterior fields at the boundary and thus the boundary field acts as a source term for the exterior theory. Furthermore, we see from \eqref{eq:path_ordered_propagator_product_state} that the path ordered propagator scales with a factor $\varepsilon^{-2}$, and it is thus not clear how this expression behaves in the continuum limit.

In order to illuminate on this issue (and to prepare for the discussion in Section \ref{sec:Beyond_the_Product_Ansatz}), we now assume that the exterior initial state $\hat{\rho}_2 (0)$ is a thermal state with respect to the exterior Hamiltonian $\hat{H}_2$, i.e., $\hat{\rho}_2 (0) = \hat{\sigma}_2 (\beta) \coloneqq Z_2^{-1} \E^{- \beta \hat{H}_2}$. Physically, this corresponds to a system where the interior and exterior fields are decoupled by imposing Dirichlet boundary conditions on the boundary and letting the exterior equilibrate to a thermal state of inverse temperature $\beta$. In this case, the expectation values in \eqref{eq:path_ordered_propagator_product_state} are taken with respect to the thermal state $\hat{\sigma}_2 (\beta)$. The influence functional in this case reads
\begin{equation}\label{eq:influence_functional_product_state_thermal}
    \begin{split}
        &\mathfrak{F}_\mathrm{IF} (\varphi^1, \varphi^2) = \int_{\mathscr{C}_\beta} \mathcal{D} \phi \; \exp \Big[ \I S_2 (\phi^1) + \I S_{12} (\varphi^1, \phi^1) \\
        &\hspace*{3em} - \I S_2 (\phi^2) - \I S_{12} (\varphi^2, \phi^2) - S^2_{\mathrm{E}} (\phi^\beta) \Big] \; ,
    \end{split}
\end{equation}
where we used the expression of thermal states in terms of imaginary time path integrals, see Section \ref{sec:Beyond_the_Product_Ansatz}.

We introduce the following $2$-point functions,
\begin{subequations}
    \begin{align}
        G^{11}_{xy} (\tau,\tau') &\coloneqq \braket{\mathcal{T} \phi_x (\tau) \, \phi_y (\tau')}_\beta \; , \\
        G^{12}_{xy} (\tau,\tau') &\coloneqq \braket{\phi_y (\tau') \, \phi_x (\tau)}_\beta \; , \\
        G^{21}_{xy} (\tau,\tau') &\coloneqq \braket{\phi_x (\tau) \, \phi_y (\tau')}_\beta \; , \\
        G^{22}_{xy} (\tau,\tau') &\coloneqq \braket{\widetilde{\mathcal{T}} \phi_x (\tau) \, \phi_y (\tau')}_\beta \; ,
    \end{align}
\end{subequations}
for $x,y \in \Lambda_2$, where $\braket{\cdot}_\beta$ denotes the expectation value with respect to the state $\hat{\sigma}_2 (\beta)$. Notice that these correlators are thermal $2$-point functions for a theory with homogeneous spatial Dirichlet boundary conditions, and we can again naturally extend them to $\overline{\Lambda_2}$ by setting $G^{ij}_{xy} (\tau,\tau') = 0$ when at least one spatial index lies on the boundary $\partial \Lambda$.

With these definitions, the path ordered propagator $\mathcal{K}$ can be written as
\begin{equation}
    \mathcal{K}_{xy} (\tau, \tau') \coloneqq \begin{pmatrix}
        g^{11}_{xy} (\tau, \tau') & - g^{12}_{xy} (\tau, \tau') \\
        - g^{21}_{xy} (\tau, \tau') & g^{22}_{xy} (\tau, \tau')
    \end{pmatrix}
\end{equation}
where we introduced the notation
\begin{equation}\label{eq:notation_double_derivative}
    g^{ij}_{xy} (\tau, \tau') \coloneqq (\partial_{n_x}^\varepsilon \partial_{n_y}^\varepsilon G^{ij})_{xy} (\tau, \tau')
\end{equation}
and $\partial_n^\varepsilon$ is again the discrete normal derivative, cf. Section \ref{sec:Lattice_Model}. We furthermore have
\begin{subequations}
    \begin{align}
        g^{11}_{xy} (\tau,\tau') &\coloneqq \braket{\mathcal{T} \partial_{n_x}^\varepsilon \phi_x (\tau) \, \partial_{n_y}^\varepsilon \phi_y (\tau')}_\beta \; , \label{eq:g11} \\
        g^{12}_{xy} (\tau,\tau') &\coloneqq \braket{\partial_{n_y}^\varepsilon \phi_y (\tau') \, \partial_{n_x}^\varepsilon \phi_x (\tau)}_\beta \; , \\
        g^{21}_{xy} (\tau,\tau') &\coloneqq \braket{\partial_{n_x}^\varepsilon \phi_x (\tau) \, \partial_{n_y}^\varepsilon \phi_y (\tau')}_\beta \; , \\
        g^{22}_{xy} (\tau,\tau') &\coloneqq \braket{\widetilde{\mathcal{T}} \partial_{n_x}^\varepsilon \phi_x (\tau) \, \partial_{n_y}^\varepsilon \phi_y (\tau')}_\beta \; , \label{eq:g22}
    \end{align}
\end{subequations}
i.e., the functions $g^{ij}_{xy}$ are the thermal $2$-point functions of the spatial normal derivatives of the field at the boundary.

Upon introducing the notation $\mathrm{x} = (\tau, x)$ and $\mathrm{y} = (\tau', y)$, we can write the continuum version of the influence action as
\begin{equation}\label{eq:influence_action_product_state_continuum}
    \begin{split}
        &S_\mathrm{IF} [\varphi^1, \varphi^2, t] = \\
        &\frac{\I}{2} \int_0^t \D \tau' \, \D \tau \int_{\partial \Omega} \D S (x) \, \D S (y) \; \varphi^a (\mathrm{x}) \, \mathcal{K}_{ab} (\mathrm{x}, \mathrm{y}) \, \varphi^b (\mathrm{y}) \; ,
    \end{split}
\end{equation}
where the kernel $\mathcal{K}$ is given by
\begin{equation}
    \mathcal{K} (\mathrm{x}, \mathrm{y}) \coloneqq \begin{pmatrix}
        g^{11} (\mathrm{x}, \mathrm{y}) & - g^{12} (\mathrm{x}, \mathrm{y}) \\
        - g^{21} (\mathrm{x}, \mathrm{y}) & g^{22} (\mathrm{x}, \mathrm{y})
    \end{pmatrix}
\end{equation}
and
\begin{equation}
    g^{ij} (\mathrm{x}, \mathrm{y}) \coloneqq \left( \frac{\partial^2}{\partial n_x \partial n_y} G^{ij} \right) (\mathrm{x}, \mathrm{y}) \; ,
\end{equation}
or more explicitly,
\begin{subequations}
    \begin{align}
        g^{11} (\mathrm{x}, \mathrm{y}) &\coloneqq \left\langle \mathcal{T} \; \frac{\partial \phi (\mathrm{x})}{\partial n_x} \, \frac{\partial \phi (\mathrm{y})}{\partial n_y} \right\rangle_\beta \; , \\
        g^{12} (\mathrm{x}, \mathrm{y}) &\coloneqq \left\langle \frac{\partial \phi (\mathrm{y})}{\partial n_y} \, \frac{\partial \phi (\mathrm{x})}{\partial n_x} \right\rangle_\beta \; , \\
        g^{21} (\mathrm{x}, \mathrm{y}) &\coloneqq \left\langle \frac{\partial \phi (\mathrm{x})}{\partial n_x} \, \frac{\partial \phi (\mathrm{y})}{\partial n_y} \right\rangle_\beta \; , \\
        g^{22} (\mathrm{x}, \mathrm{y}) &\coloneqq \left\langle \widetilde{\mathcal{T}} \; \frac{\partial \phi (\mathrm{x})}{\partial n_x} \, \frac{\partial \phi (\mathrm{y})}{\partial n_y} \right\rangle_\beta \; .
    \end{align}
\end{subequations}

In summary, the influence action $S_\mathrm{IF}$ for the product ansatz \eqref{eq:product_state} is a bilinear form in the fields $\varphi^1$ and $\varphi^2$ supported on the spatial boundary $\partial \Omega$, i.e., it is a boundary action. The kernel of this bilinear form is given by a path ordered propagator $\mathcal{K}$ of the exterior theory, which is a $2 \times 2$ matrix whose entries are given by the functions $g^{ij}$, $i,j \in \{1,2\}$. These functions are the thermal $2$-point functions of the spatial normal derivatives of the exterior field at the boundary. Alternatively, by linearity, the functions $g^{ij}$ are the double spatial normal derivatives of the thermal $2$-point functions $G^{ij}$ of the exterior field at the boundary. The functions $G^{ij}$ are the usual (thermal) Schwinger-Keldysh $2$-point functions, however not for a theory on $\R^D$ but for a theory on the exterior region with homogeneous spatial Dirichlet boundary conditions. We stress once again that in the above considerations, the exterior field $\phi$ is regarded as a field with Dirichlet boundary conditions on the boundary $\partial \Omega$, rather than as the restriction of a global field to the exterior region.

\subsection{Initial State Correlations}\label{sec:Beyond_the_Product_Ansatz}
In this Section we generalize the result from the previous Section to the case where the initial state of the system is not a product state, i.e., it contains correlations between the system and the environment \cite{Hakim1985,Smith1987,Romero1997}. More precisely, following, e.g., \cite{Romero1997}, we consider an initial state of the form
\begin{equation}
    \begin{split}
        &\rho_{\beta,\lambda} (0; \varphi^1, \phi^1, \varphi^2, \phi^2) = \\
        &\int [\D \overline{\varphi}^1] [\D \overline{\varphi}^2] \; \lambda (\varphi^1, \overline{\varphi}^1, \varphi^2, \overline{\varphi}^2) \, \rho_{\beta} (0; \overline{\varphi}^1, \phi^1, \overline{\varphi}^2, \phi^2) \; ,
    \end{split}
\end{equation}
where we defined
\begin{equation}
    [\D \overline{\varphi}^i] \coloneqq \prod_{x \in \Lambda_1} \D \overline{\varphi}_x^i \; .
\end{equation}
Here, $\rho_{\beta}$ is the density kernel of the canonical thermal state at inverse temperature $\beta$ and $\lambda$, called the preparation function in \cite{Romero1997}, parametrizes a deviation from the thermal state within the region $\Lambda_1$. We assume that the initial thermal state $\rho_{\beta}$ is Gaussian (which amounts to the requirement that the Hamiltonian is quadratic, i.e., $V \equiv 0$ in \eqref{eq:Hamiltonian_operator_anharmonic}).

Let us now explain the preparation function $\lambda$ in more detail. The preparation function $\lambda$ is used to incorporate non-equilibrium states in the form of perturbations of the thermal state localized in the interior region. As a limiting case, one can consider the choice
\begin{equation}
    \lambda (\varphi^1, \overline{\varphi}^1, \varphi^2, \overline{\varphi}^2) = \delta (\varphi^1 - \overline{\varphi}^1) \, \delta (\varphi^2 - \overline{\varphi}^2) \; ,
\end{equation}
in which case $\rho_{\beta,\lambda} = \rho_{\beta}$, i.e., the global initial state is the canonical thermal state and no local perturbation is present.

As a concrete example of a genuine non-equilibrium state, one can consider locally excited coherent states \cite{Ciolli2019,Longo2019,Casini2019,Bostelmann2021}. Let $\beta = + \infty$, i.e., we consider the vacuum state whose state vector is denoted by $\ket{\Omega}$. Then, the density kernel of the vacuum state can be written as
\begin{equation}
    \rho_\mathrm{vac} (\varphi^1, \phi^1, \varphi^2, \phi^2) = \braket{\varphi^1, \phi^1 | \Omega} \braket{\Omega | \varphi^2, \phi^2} \; .
\end{equation}
Let $\hat{W} (f,g)$ be the Weyl operator given by
\begin{equation}
    \hat{W} (f,g) = \exp \left[ \I \int_\mathscr{L} \left( f_x \hat{\Phi}_x + g_x \hat{\Pi}_x \right) \D^d x \right] \; ,
\end{equation}
where $f$ and $g$ are real-valued functions on the lattice $\mathscr{L}^d_{\varepsilon,L}$ supported in the interior region $\Lambda_1$. The Weyl operator $\hat{W} (f,g)$ defines a coherent state $\ket{\Omega_{f,g}}$ by acting on the vacuum state, i.e., $\ket{\Omega_{f,g}} = \hat{W} (f,g) \ket{\Omega}$. For simplicity, let $g \equiv 0$ in the following. Then, the kernel of the Weyl operator $\hat{W} (f) \coloneqq \hat{W} (f,0)$ is given by
\begin{equation}
    \begin{split}
        W_f (\varphi^1, \phi^1, \varphi^2, \phi^2) &= \braket{\varphi^1, \phi^1 | \, \hat{W} (f) \, | \varphi^2, \phi^2} \\
        &= \E^{\I \braket{\varphi^2, f}_\varepsilon} \; \delta (\varphi^1 - \varphi^2) \delta (\phi^1 - \phi^2) \; ,
    \end{split}
\end{equation}
where the inner product $\braket{\cdot,\cdot}_\varepsilon$ was defined in Section \ref{sec:Lattice_Model}. The density kernel of the coherent state $\ket{\Omega_{f}}$ is then given by
\begin{equation}
    \begin{split}
        &\rho_f (\varphi^1, \phi^1, \varphi^2, \phi^2) = \int [\D \overline{\varphi}^1] [\D \overline{\varphi}^2] \; \E^{\I \braket{\overline{\varphi}^1 - \overline{\varphi}^2, f}_\varepsilon} \\
        &\times \delta (\varphi^1 - \overline{\varphi}^1) \, \delta (\varphi^2 - \overline{\varphi}^2) \, \rho_\mathrm{vac} (\overline{\varphi}^1, \phi^1, \overline{\varphi}^2, \phi^2) \; .
    \end{split}
\end{equation}
Therefore, the preparation function $\lambda$ for the coherent state $\ket{\Omega_{f}}$ is given by
\begin{equation}
    \lambda (\varphi^1, \overline{\varphi}^1, \varphi^2, \overline{\varphi}^2) = \E^{\I \braket{\overline{\varphi}^1 - \overline{\varphi}^2, f}_\varepsilon} \, \delta (\varphi^1 - \overline{\varphi}^1) \, \delta (\varphi^2 - \overline{\varphi}^2) \; .
\end{equation}
In general, the preparation function $\lambda$ can be chosen to describe a wide range of non-equilibrium initial states, including also non-Gaussian perturbations.

The global thermal density kernel can be written as a path integral over field configurations on the imaginary time interval $- \I [0,\beta]$. More precisely, we have
\begin{equation}\label{eq:thermal_density_kernel}
    \rho_{\beta} (0; \overline{\varphi}^1, \phi^1, \overline{\varphi}^2, \phi^2) = \frac{1}{Z_\beta} \int_{\overline{\varphi}^2, \phi^2}^{\overline{\varphi}^1, \phi^1} \mathcal{D} \widetilde{\varphi} \, \mathcal{D} \widetilde{\phi} \; \E^{- S_\mathrm{E} (\widetilde{\varphi}, \widetilde{\phi})} \; ,
\end{equation}
where the Euclidean action $S_\mathrm{E}$ is given by
\begin{equation}
    \begin{split}
        &S_\mathrm{E} (\psi) = \\
        &\frac{1}{2} \int_0^\beta \D \tau \int_{\mathscr{L}} \D^d x \left[ \left( \frac{\partial \psi_x}{\partial \tau} \right)^2 + \int_{\mathscr{L}} \D^d y \, \psi_x (\tau) \, D^\varepsilon_{xy} \, \psi_y (\tau) \right] \, .
    \end{split}
\end{equation}
The integral boundaries in \eqref{eq:thermal_density_kernel} indicate that the field configurations $\widetilde{\varphi}$ and $\widetilde{\phi}$ on $- \I [0,\beta] \times \Lambda_1$ and $- \I [0,\beta] \times \Lambda_2$, respectively, are held fixed at ``times'' $0$ and $- \I \beta$ and take the prescribed values $\overline{\varphi}^2, \phi^2$ and $\overline{\varphi}^1, \phi^1$ there.
\begin{figure}[t]
    \centering
    \includegraphics[width=0.4\textwidth]{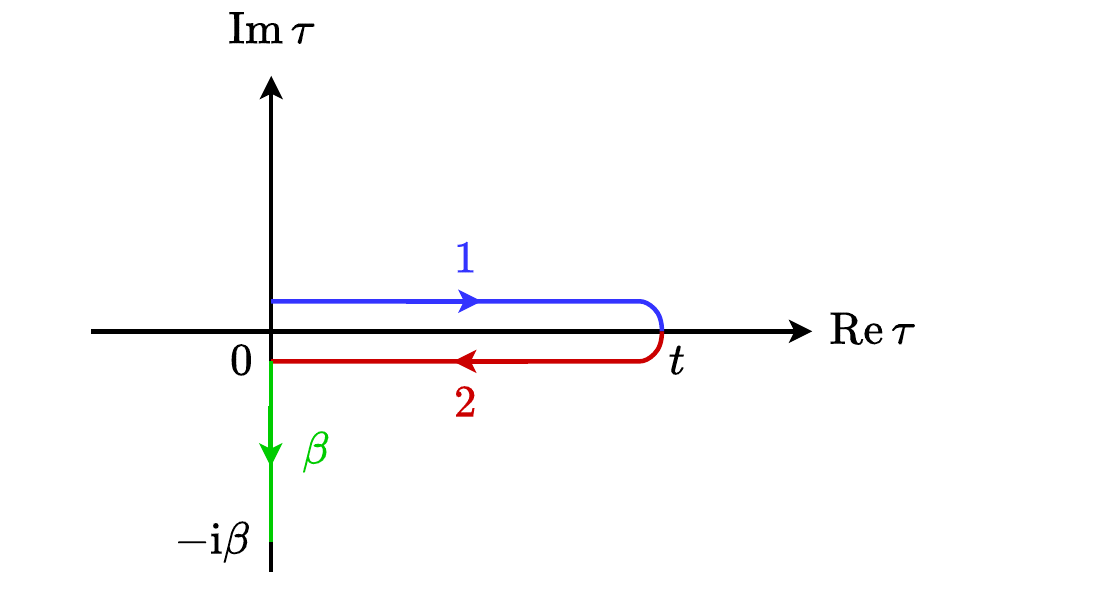}
    \caption[Schwinger-Keldysh contour $\mathscr{C}_\beta$ for the influence functional in the presence of a thermal initial state.]{Schwinger-Keldysh contour $\mathscr{C}_\beta$ for the influence functional in the presence of a thermal initial state. In addition to the forward and backward branches, the contour also contains a branch $\beta$ over the imaginary time interval from $0$ to $- \I \beta$, accounting for the thermal part of the initial state. The contour is closed, i.e., the field at imaginary time $- \I \beta$ on the $\beta$ branch is identified with the field at time $0$ on the forward branch. For illustration purposes, the real time branches are deformed into the complex plane.}
    \label{fig:SK_contour_beta}
\end{figure}

With this setup, it was shown in \cite[Sec.~III]{Romero1997} that the reduced density kernel at time $t$ can be obtained by evolving the initial preparation function, i.e.,
\begin{equation}\label{eq:reduced_density_kernel_thermal}
    \begin{split}
        &\rho_{\mathrm{S}, \beta, \lambda} (t; \varphi^1, \varphi^2) = \frac{1}{Z_\beta} \int^{\varphi^1, \varphi^2} \mathcal{D} \widetilde{\varphi}^1 \mathcal{D} \widetilde{\varphi}^2 \mathcal{D} \widetilde{\varphi}^\beta \times \\
        &\exp \left[ \I (S_1 (\widetilde{\varphi}^1) - S_1 (\widetilde{\varphi}^2)) - S^1_{\mathrm{E}} (\widetilde{\varphi}^\beta )+ \I S^\beta_\mathrm{IF} (\widetilde{\varphi}^1,\widetilde{\varphi}^2,\widetilde{\varphi}^\beta,t) \right] \times \\
        &\lambda (\varphi_0^1, \varphi_0^{\beta,1}, \varphi_0^2, \varphi_0^{\beta,2}) \; ,
    \end{split}
\end{equation}
where the \emph{generalized} influence action $S^\beta_\mathrm{IF}$ is given by the generalized Feynman-Vernon influence functional $\mathfrak{F}_\mathrm{IF}$ as
\begin{equation}\label{eq:influence_functional_thermal}
    \begin{split}
        &\mathfrak{F}^\beta_\mathrm{IF} (\varphi^1, \varphi^2, \varphi^\beta) = \E^{\I S^\beta_\mathrm{IF} (\varphi^1, \varphi^2, \varphi^\beta,t)} \\
        &= \int_{\mathscr{C}_\beta} \mathcal{D} \phi \; \exp \Big[ \I S_2 (\phi^1) + \I S_{12} (\varphi^1, \phi^1) - \I S_2 (\phi^2) \\
        &\hspace*{3em} - \I S_{12} (\varphi^2, \phi^2) - S^2_{\mathrm{E}} (\phi^\beta) - S^{12}_{\mathrm{E}} (\varphi^\beta, \phi^\beta )\Big] \; ,
    \end{split}
\end{equation}
where $\mathscr{C}_\beta$ is the closed Schwinger-Keldysh contour over two connected time paths as well as over an imaginary time contour accounting for the thermal part of the initial state as shown in Fig. \ref{fig:SK_contour_beta}. Again, the upper limit of the path integral in \eqref{eq:reduced_density_kernel_thermal} indicates that the fields $\widetilde{\varphi}^1$ and $\widetilde{\varphi}^2$ on the forward and backward branches, respectively, are held fixed at time $t$ and take the prescribed values $\varphi^1$ and $\varphi^2$ there.

The above expression for the generalized influence functional should be compared to the influence functional in \eqref{eq:influence_functional_product_state_thermal} for the case of a product state with thermal environment. The only difference is the additional term $S^{12}_{\mathrm{E}}$ in the exponent, which acts as a boundary source term on the thermal branch. Therefore, we conclude that the initial state correlations are incorporated by a linear system-environment coupling across the spatial boundary on the thermal branch and the expression for the influence functional may be seen as a Gaussian integral over the contour $\mathscr{C}_\beta$ with a (both real and imaginary) time dependent source term on the boundary.

Just like in the last Section, this integral can be solved explicitly and yields the following expression for the influence action
\begin{equation}
    \begin{split}
        &S^\beta_\mathrm{IF} (\varphi^1, \varphi^2, \varphi^\beta, t) = \\
        &\frac{\I}{2} \int_{I^{a,b}} \D \tau \D \tau' \int_{\partial \Lambda} \D S (x) \D S (y) \, \varphi^a_x (\tau) ( \mathcal{K}^\beta_{xy} )_{ab} (\tau, \tau') \varphi^b_y (\tau') \; ,
    \end{split}
\end{equation}
where summation over time-path indices $a,b \in \{ 1,2,\beta \}$ is assumed and the integration interval $I^a$ is given by $I^1 = I^2 = (0,t)$ and $I^\beta = (0,\beta)$. The kernel $\mathcal{K}^\beta$ is given by
\begin{equation}
    \mathcal{K}^\beta_{xy} (\tau, \tau') \coloneqq \begin{pmatrix}
        g^{11}_{xy} (\tau, \tau') & - g^{12}_{xy} (\tau, \tau') & \I g^{1 \beta}_{xy} (\tau, \tau') \\
        - g^{21}_{xy} (\tau, \tau') & g^{22}_{xy} (\tau, \tau') & - \I g^{2 \beta}_{xy} (\tau, \tau') \\
        \I g^{\beta 1}_{xy} (\tau, \tau') & - \I g^{\beta 2}_{xy} (\tau, \tau') & - g^{\beta \beta}_{xy} (\tau, \tau')
    \end{pmatrix} \; ,
\end{equation}
where again a lower case $g$ indicates a double (discrete) normal derivative in the spatial indices, cf. \eqref{eq:notation_double_derivative} and \eqref{eq:g11} -- \eqref{eq:g22}, and we introduced the new $2$-point functions
\begin{subequations}
    \begin{align}
        \begin{split}
            G^{\beta \beta}_{xy} (t,t') &\coloneqq \braket{\phi_x (- \I t) \, \phi_y (- \I t')}_\beta \\
            &\hphantom{:} = \braket{\phi_y (- \I t') \, \phi_x (- \I t)}_\beta \; ,
        \end{split} \\
        G^{i \beta}_{xy} (t,t') &\coloneqq \braket{\phi_y (- \I t') \, \phi_x (t)}_\beta \; , \\
        G^{\beta i}_{xy} (t,t') &\coloneqq \braket{\phi_x (- \I t) \, \phi_y (t')}_\beta \; ,
    \end{align}
\end{subequations}
for $x,y \in \Lambda_2$ and $i \in \{ 1,2 \}$. Here, the expectation values are again the thermal expectation values with respect to the exterior state $\hat{\sigma}_2 (\beta)$ as indicated by the subscript $\beta$. Notice that $G^{\beta 1} = G^{\beta 2}$ and $G^{1 \beta} = G^{2 \beta}$. In continuum notation, the influence action can be written as
\begin{equation}
    \begin{split}
        &S^\beta_\mathrm{IF} [\varphi^1, \varphi^2, \varphi^\beta, t] = \\
        &\frac{\I}{2} \int_{I^{a,b}} \D \tau \, \D \tau' \int_{\partial \Omega} \D S (x) \, \D S (y) \; \varphi^a (\mathrm{x}) \, \mathcal{K}^\beta_{ab} (\mathrm{x}, \mathrm{y}) \, \varphi^b (\mathrm{y}) \; ,
    \end{split}
\end{equation}
where the kernel $\mathcal{K}^\beta$ is given by
\begin{equation}\label{eq:kernel_generalized_influence_action}
    \mathcal{K}^\beta (\mathrm{x}, \mathrm{y}) = \begin{pmatrix}
        g^{11} (\mathrm{x}, \mathrm{y}) & - g^{12} (\mathrm{x}, \mathrm{y}) & \I g^{1 \beta} (\mathrm{x}, \mathrm{y}) \\
        - g^{21} (\mathrm{x}, \mathrm{y}) & g^{22} (\mathrm{x}, \mathrm{y}) & - \I g^{2 \beta} (\mathrm{x}, \mathrm{y}) \\
        \I g^{\beta 1} (\mathrm{x}, \mathrm{y}) & - \I g^{\beta 2} (\mathrm{x}, \mathrm{y}) & - g^{\beta \beta} (\mathrm{x}, \mathrm{y})
    \end{pmatrix} \; .
\end{equation}

We note that the generalized influence action $S^\beta_\mathrm{IF}$ can be written as
\begin{equation}
    \begin{split}
        S^\beta_\mathrm{IF} (\varphi^1, \varphi^2, \varphi^\beta, t) &= S_\mathrm{IF} (\varphi^1, \varphi^2, t) + S^\mathrm{corr.}_\mathrm{IF} (\varphi^1, \varphi^2, \varphi^\beta, t) \\
        &- \I \Delta S^1_\mathrm{E} (\varphi^\beta) \; ,
    \end{split}
\end{equation}
where the first term is the influence action for the product ansatz with thermal environment (see Section \ref{sec:Factorizing_Initial_State}), the second term accounts for the initial correlations between the system and the environment, and the third term contains the non-local boundary conditions for the Euclidean action of the region $\Lambda_1$, cf. Section \ref{sec:Euclidean_Field_Theory}. More precisely, the third term is given by
\begin{equation}
    \begin{split}
        &\Delta S^1_\mathrm{E} (\varphi^\beta) = \\
        &\frac{1}{2} \int_0^\beta \D \tau \, \D \tau' \int_{\partial \Lambda} \D S (x) \, \D S (y) \; \varphi_x^\beta (\tau) \, g_{xy}^{\beta \beta} (\tau, \tau') \, \varphi_y^\beta (\tau') \; ,
    \end{split}
\end{equation}
and in particular we have
\begin{equation}
    \widetilde{S^1_\mathrm{E}} (\varphi^\beta) = S^1_\mathrm{E} (\varphi^\beta) - \Delta S^1_{\mathrm{E}} (\varphi^\beta) \; ,
\end{equation}
see \eqref{eq:tilde_euclidean_action} and the related discussion in Section \ref{sec:Euclidean_Field_Theory}. Note that the term $S^\mathrm{corr.}_\mathrm{IF}$ accounts for the contribution of the initial state correlations to the dynamics. It stems from the $1 \beta$, $2 \beta$, $\beta 1$ and $\beta 2$ entries of the kernel $\mathcal{K}^\beta$ in \eqref{eq:kernel_generalized_influence_action}.

Plugging these results back into the expression for the reduced density kernel, we obtain
\begin{equation}
    \begin{split}
        &\rho_{\mathrm{S}, \beta, \lambda} (t; \varphi^1, \varphi^2) = \\
        &\int^{\varphi^1, \varphi^2} \mathcal{D} \widetilde{\varphi}^1 \mathcal{D} \widetilde{\varphi}^2 \mathcal{D} \widetilde{\varphi}^\beta \; \E^{\I S^\beta_\mathrm{eff} (\widetilde{\varphi}^1, \widetilde{\varphi}^2, \widetilde{\varphi}^\beta, t) - \widetilde{S^1_\mathrm{E}} (\widetilde{\varphi}^\beta)} \times \\
        &\hspace*{1em} \lambda (\varphi_0^1, \varphi_0^{\beta,1}, \varphi_0^2, \varphi_0^{\beta,2}) \; ,
    \end{split}
\end{equation}
where we defined the effective action $S^\beta_\mathrm{eff}$ as
\begin{equation}\label{eq:effective_action_initial_correlations}
    \begin{split}
        S^\beta_\mathrm{eff} (\varphi^1, \varphi^2, \varphi^\beta, t) &\coloneqq S_1 (\varphi^1) - S_1 (\varphi^2) + S_\mathrm{IF} (\varphi^1, \varphi^2, t ) \\
        &+ S^\mathrm{corr.}_\mathrm{IF} (\varphi^1, \varphi^2, \varphi^\beta, t) \; .
    \end{split}
\end{equation}
Notice that the limit $t \to 0^+$ reproduces the correct initial condition for the system.

We remark that the effect of initial correlations between the system and the environment is again encoded in boundary terms, yielding non-local boundary conditions for the differential operator in the action.

\subsection{Dynamics of Field Expectation Values}\label{sec:Dynamics_of_Field_Expectation_Values}

In this Section, we derive a stochastic equation of motion for the expectation values of the field in the interior region. The model we consider is a non-interacting continuum field theory with a locally perturbed thermal initial state, see Section \ref{sec:Beyond_the_Product_Ansatz}, which is assumed to be Gaussian. The equation of motion for the expectation values of the field in the interior region is then derived by taking the functional derivative of the effective action with respect to the field.

\subsubsection{Derivation of the Stochastic Equation of Motion}\label{sec:Derivation_of_the_Stochastic_Equation_of_Motion}

The closed time path generating functional $W_\lambda [J^1, J^2]$ of connected correlation functions of a locally perturbed thermal state described by a preparation function $\lambda$ (see Section \ref{sec:Beyond_the_Product_Ansatz}) is given by \cite[Sec.~6.3]{Calzetta2022}
\begin{equation}
    \E^{\I W_\lambda [J^1, J^2]} = \mathrm{Tr} \left\lbrace \hat{\rho}_{\mathrm{S},\beta,\lambda} \; \E^{\I \int_{\mathscr{C}, \Omega} J \varphi} \right\rbrace \; .
\end{equation}
For simplicity, we will for the rest of this Section assume that the initial reduced state $\hat{\rho}_{\mathrm{S},\beta,\lambda}$ is Gaussian.

The closed time path quantum effective action (CTPQEA) $\Gamma_\lambda [\Phi^1, \Phi^2]$ is defined as the functional Legendre transform of the CTP generating functional $W_\lambda [J^1, J^2]$ with respect to the sources $J^1$ and $J^2$, i.e.,
\begin{equation}
    \Gamma_\lambda [\Phi^1, \Phi^2] = \sup_{J^1, J^2} \left\lbrace W_\lambda [J^1, J^2] - \int_{\mathscr{C}, \Omega} J \Phi \right\rbrace \; ,
\end{equation}
where the average field $\Phi^i$ is given by
\begin{equation}
    \Phi^i (\mathrm{x}) = \frac{\delta W_\lambda [J^1, J^2]}{\delta J^i (\mathrm{x})}
\end{equation}
and $\Phi^1 = \Phi^2$ if $J^1 = J^2 = 0$.

The quantum equation of motion for the expectation value of the field in the interior region is derived from the CTPQEA via the principle of stationary action. In the case of a purely Gaussian theory with perturbed thermal initial state correlations as described in Section \ref{sec:Beyond_the_Product_Ansatz} (recall that we assume $\hat{\rho}_{\mathrm{S},\beta,\lambda}$ to be Gaussian), the CTPQEA reads
\begin{equation}
    \Gamma_\lambda [\Phi^1, \Phi^2] = S_1 [\Phi^1] - S_1 [\Phi^2] + S_\mathrm{IF} [\Phi^1, \Phi^2, t] + F_\lambda [\Phi_0^1, \Phi_0^2] \; ,
\end{equation}
where $F_\lambda$ is a bilinear form in the $t=0$ fields $\Phi_0^1$ and $\Phi_0^2$, accounting for the initial state $\hat{\rho}_{\mathrm{S},\beta,\lambda}$.

It is convenient to introduce, following \cite{Calzetta2022}, the difference field $\Psi \coloneqq \Phi^1 - \Phi^2$ and the ``centre of mass'' field $\Xi \coloneqq \frac{1}{2} (\Phi^1 + \Phi^2)$. Using the difference and centre of mass fields, the influence action $S_\mathrm{IF}$ of the product ansatz \eqref{eq:influence_action_product_state_continuum} can be written as \cite[Sec.~3.2.2]{Calzetta2022}
\begin{equation}\label{eq:influence_action_dissipation_noise}
    \begin{split}
        &S_\mathrm{IF} [\Psi, \Xi, t] = \int_0^t \D \tau \, \D \tau' \int_{\partial \Omega} \D S (x) \, \D S (y) \times \\
        &\hspace*{2em} \left[ \Psi (\mathrm{x}) \, \boldsymbol{\mathrm{D}} (\mathrm{x}, \mathrm{y}) \, \Xi (\mathrm{y}) + \frac{\I}{2} \Psi (\mathrm{x}) \, \boldsymbol{\mathrm{N}} (\mathrm{x}, \mathrm{y}) \, \Psi (\mathrm{y}) \right] \; ,
    \end{split}
\end{equation}
where $\boldsymbol{\mathrm{D}}$ and $\boldsymbol{\mathrm{N}}$ are the dissipation and noise kernels, respectively, given by the boundary terms
\begin{align}
    \boldsymbol{\mathrm{D}} (\mathrm{x}, \mathrm{y}) &= \I \Theta (\tau - \tau') \left( g^{21} (\mathrm{x}, \mathrm{y}) - g^{12} (\mathrm{x}, \mathrm{y}) \right) \; , \label{eq:dissipation_kernel} \\
    \boldsymbol{\mathrm{N}} (\mathrm{x}, \mathrm{y}) &= \frac{1}{2} \left( g^{21} (\mathrm{x}, \mathrm{y}) + g^{12} (\mathrm{x}, \mathrm{y}) \right) \; . \label{eq:noise_kernel}
\end{align}
In the following, we will denote by $\hat{\boldsymbol{\mathrm{D}}}$ and $\hat{\boldsymbol{\mathrm{N}}}$ the integral operators with kernels $\boldsymbol{\mathrm{D}} (\mathrm{x}, \mathrm{y})$ and $\boldsymbol{\mathrm{N}} (\mathrm{x}, \mathrm{y})$, respectively.

Before proceeding, we discuss the dissipation and noise kernels in \eqref{eq:dissipation_kernel} and \eqref{eq:noise_kernel}, respectively, in the operator picture. Notice that the dissipation kernel $\boldsymbol{\mathrm{D}}$ and the noise kernel $\boldsymbol{\mathrm{N}}$ can be written as
\begin{align}
    \boldsymbol{\mathrm{D}} (\mathrm{x}, \mathrm{y}) &= \I \Theta (\tau - \tau') \; \frac{\partial^2}{\partial n_x \partial n_y} \left\langle \left[ \hat{\Phi}_\mathrm{D} (\mathrm{x}), \hat{\Phi}_\mathrm{D} (\mathrm{y}) \right] \right\rangle_\beta \; , \label{eq:dissipation_kernel_operator} \\
    \boldsymbol{\mathrm{N}} (\mathrm{x}, \mathrm{y}) &= \frac{1}{2} \frac{\partial^2}{\partial n_x \partial n_y} \left\langle \left\lbrace \hat{\Phi}_\mathrm{D} (\mathrm{x}), \hat{\Phi}_\mathrm{D} (\mathrm{y}) \right\rbrace \right\rangle_\beta \; , \label{eq:noise_kernel_operator}
\end{align}
where $[.,.]$ and $\{.,.\}$ denote the commutator and anti-commutator, respectively, and the expectation values are taken with respect to the thermal state $\hat{\sigma}_\beta$ of the exterior field operator $\hat{\Phi}_\mathrm{D}$. The subscript $\mathrm{D}$ of the field operator indicates that the exterior field theory is quantized with Dirichlet boundary conditions on $\partial \Omega$. In particular, this means that
\begin{equation}
    \left[ \hat{\Phi}_\mathrm{D} (\mathrm{x}), \hat{\Phi}_\mathrm{D} (\mathrm{y}) \right] \neq \I \Delta_\mathrm{PJ} (\mathrm{x} - \mathrm{y}) \, \mathds{1} \; ,
\end{equation}
where $\Delta_\mathrm{PJ}$ is the Pauli-Jordan distribution or commutator function \cite{Bogolyubov1959,Bjorken1965}. Rather, the commutator function is modified in order to incorporate Dirichlet boundary conditions on the spatial boundary $\partial \Omega$.

We observe that the dissipation kernel $\boldsymbol{\mathrm{D}}$ has the structure of a double normal derivative of the retarded propagator of the exterior theory, which can equivalently be interpreted as a response function, see Section \ref{sec:Linear_Response_Theory}. In contrast, the noise kernel is defined as the double normal derivative of the quantum correlation function of the exterior field. These are the expected structures for the dissipation and noise kernels, respectively, see, e.g., \cite{Calzetta2022}. However, by the local nature of the theory considered in this work, all quantities are restricted to the spatial boundary $\partial \Omega$.

Upon writing $S_1 [\varphi^1] - S_1 [\varphi^2]$ in terms of the fields $\Psi$ and $\Xi$, we arrive at the expression
\begin{equation}
    S_1 [\varphi^1] - S_1 [\varphi^2] = \int_{t, \Omega} \left[ \dot{\Psi} \dot{\Xi} - \nabla \Psi \cdot \nabla \Xi - m^2 \Psi \Xi \right] \; ,
\end{equation}
where we introduced the notation
\begin{align}
    \int_{t, \Omega} &\coloneqq \int_0^t \D \tau \int_{\Omega} \D^d x \; , \\
    \int_{t, \partial \Omega} &\coloneqq \int_0^t \D \tau \int_{\partial \Omega} \D S (x) \; .
\end{align}

In summary, the effective action $S_\mathrm{eff} [\Phi^1, \Phi^2, t] \coloneqq S_1 [\Phi^1] - S_1 [\Phi^2] + S_\mathrm{IF} [\Phi^1, \Phi^2, t]$ can be written in terms of the fields $\Psi$ and $\Xi$ as
\begin{equation}
    \begin{split}
        &S_\mathrm{eff} [\Psi, \Xi, t] = \int_{t, \Omega} \left[ \dot{\Psi} \dot{\Xi} - \nabla \Psi \cdot \nabla \Xi - m^2 \Psi \Xi \right] \\
        &+ \iint_{t, \partial \Omega} \left[ \Psi (\mathrm{x}) \boldsymbol{\mathrm{D}} (\mathrm{x}, \mathrm{y}) \, \Xi (\mathrm{y}) + \frac{\I}{2} \Psi (\mathrm{x}) \, \boldsymbol{\mathrm{N}} (\mathrm{x}, \mathrm{y}) \, \Psi (\mathrm{y}) \right] \; .
    \end{split}
\end{equation}
We furthermore can write the factor containing the noise kernel as the Fourier transform (or characteristic functional) of a Gaussian measure, i.e.,
\begin{equation}
    \begin{split}
        &\exp \left[ - \frac{1}{2} \iint_{t, \partial \Omega} \Psi (\mathrm{x}) \, \boldsymbol{\mathrm{N}} (\mathrm{x}, \mathrm{y}) \, \Psi (\mathrm{y}) \right] = \\
        &\hspace*{5em} \int \D \mathcal{P} (\xi) \; \exp \left[ \I \int_{t, \partial \Omega} \xi (\mathrm{x}) \Psi (\mathrm{x}) \right] \; ,
    \end{split}
\end{equation}
where $\mathcal{P}$ is a centred Gaussian measure with covariance given by
\begin{equation}
    \mathbb{E}_\mathcal{P} [\xi (\mathrm{x}) \xi (\mathrm{y})] = \boldsymbol{\mathrm{N}} (\mathrm{x}, \mathrm{y}) \; .
\end{equation}
Thus, we can write
\begin{equation}
    \begin{split}
        &\E^{\I S_\mathrm{eff.} [\Psi, \Xi, t]} = \\
        &\int \D \mathcal{P} (\xi) \; \exp \bigg[ \I \left( \int_{t, \Omega} \dot{\Psi} \, \dot{\Xi} - \nabla \Psi \cdot \nabla \Xi - m^2 \Psi \Xi \right) \\
        &+ \I \left( \int_{t, \partial \Omega} \xi \Psi + \Psi \hat{\boldsymbol{\mathrm{D}}} \Xi \right) \bigg] \; .
    \end{split}
\end{equation}

We may now write the effective action as containing the Gaussian random variable $\xi$ as
\begin{equation}
    \begin{split}
        S^\xi_\mathrm{eff.} [\Psi, \Xi, t] &= \int_{t, \Omega} \left( \dot{\Psi} \, \dot{\Xi} - \nabla \Psi \cdot \nabla \Xi - m^2 \Psi \Xi \right) \\
        &+ \int_{t, \partial \Omega} \left( \xi \Psi + \Psi \hat{\boldsymbol{\mathrm{D}}} \Xi \right) \; .
    \end{split}
\end{equation}
For the above expression to make sense, we need to keep in mind that ultimately we need to average with respect to the measure $\mathcal{P}$.

The stochastic equation of motion for the expectation value of the interior field is obtained by extremizing the effective action with respect to the difference field $\Psi$, i.e. (see \cite[Sec.~5.1.3]{Calzetta2022} and \cite[App.~A]{Greiner1997}),
\begin{equation}
    \left. \frac{\delta S^\xi_\mathrm{eff.} [\Psi, \Xi, t]}{\delta \Psi} \right|_{\Psi = 0} = 0 \; .
\end{equation}
For the case at hand and upon denoting the expectation value of the field $\varphi$ by $\Phi$, this yields the following stochastic initial-boundary value problem for the Klein-Gordon equation,
\begin{subequations}
    \begin{alignat}{2}
        &\left( \Box + m^2 \right) \Phi (\mathrm{x}) = 0 \; , \quad\;\; &&\mathrm{x} \in (0,t] \times \Omega \; , \label{eq:stochastic_Klein_Gordon} \\
        &\frac{\partial \Phi}{\partial n} (\mathrm{x}) + (\hat{\boldsymbol{\mathrm{D}}} \Phi) (\mathrm{x}) = - \xi (\mathrm{x}) \; , \quad\;\; &&\mathrm{x} \in [0,t] \times \partial \Omega \; , \label{eq:stochastic_boundary_condition} \\
        &\begin{cases}
            \Phi (0, x) = f (x) \\
            \dot{\Phi} (0,x) = g (x)
        \end{cases} , \quad\;\; &&x \in \Omega \; , \label{eq:initial_conditions}
    \end{alignat}
\end{subequations}
where $\Box \coloneqq \frac{\partial^2}{\partial \tau^2} - \Delta$ is the d'Alembert operator and $f$ and $g$ are the initial conditions for $\Phi$ and its time derivative, respectively.

We observe that the stochastic part of the dynamics is encoded in the spatial boundary conditions for the field. The noise kernel $\boldsymbol{\mathrm{N}}$ induces a stochastic ``force'' term $- \xi$ in \eqref{eq:stochastic_boundary_condition} which is dynamical in nature, i.e., it stems from the linear system-environment coupling in the classical action. The dissipation kernel $\boldsymbol{\mathrm{D}}$ causes the boundary conditions to be non-local both in space and time. More precisely, fix some $\mathrm{x} = (t', x) \in (0,t) \times \partial \Omega$. Then, the term $(\hat{\boldsymbol{\mathrm{D}}} \Phi) (\mathrm{x})$ in \eqref{eq:stochastic_boundary_condition} is given by
\begin{equation}\label{eq:Dv}
    \begin{split}
        &(\hat{\boldsymbol{\mathrm{D}}} \Phi) (\mathrm{x}) = \int_0^t \D \tau' \int_{\partial \Omega} \D S (y) \; \boldsymbol{\mathrm{D}} (\mathrm{x}, \mathrm{y}) \, \Phi (\mathrm{y}) \\
        &= \I \int_0^{t'} \D \tau' \int_{\partial \Omega} \D S (y) \, \left( g^{21} - g^{12} \right) (\mathrm{x}, \mathrm{y}) \; \Phi (\mathrm{y}) \; .
    \end{split}
\end{equation}
Notice that the Heaviside function in the expression for the dissipation kernel $\boldsymbol{\mathrm{D}}$ (see \eqref{eq:dissipation_kernel}) restricts the time integral to the interval $[0,t']$. We see that the normal derivative of the field expectation value $v$ at a point $\mathrm{x} = (t',x)$ of the spacetime boundary $\R_+ \times \partial \Omega$ depends on the value of the field expectation value at all points of the spatial boundary $\partial \Omega$ at all times up to $t'$. This generalizes the result from Section \ref{sec:Euclidean_Field_Theory}, where the normal derivative depends on the value of the field on the whole boundary $\partial \Omega$ of some Euclidean region $\Omega$. The presence of a memory integral in \eqref{eq:Dv}, i.e., the dependence of the normal derivative on the value of the field expectation in the whole past, indicates that the time evolution described by \eqref{eq:stochastic_Klein_Gordon} -- \eqref{eq:initial_conditions} is non-Markovian.

\subsubsection{Linear Response Theory}\label{sec:Linear_Response_Theory}

We now want to discuss the results obtained in this Section in view of linear response theory \cite{Kubo1991,Altland2010}. Notice that the dissipation kernel $\boldsymbol{\mathrm{D}}$ given in \eqref{eq:dissipation_kernel} is the retarded Green's function of the spatial normal derivatives of the fields on the boundary in the exterior Dirichlet theory in a canonical thermal state of inverse temperature $\beta$. Since the retarded propagator describes the response of a system to an external perturbation, it is also called response function.

Recall the expression for the influence functional in \eqref{eq:influence_functional_thermal}.  We can write it compactly and in continuum notation as
\begin{equation}
    \begin{split}
        &\mathfrak{F}^\beta_\mathrm{IF} [\varphi^1, \varphi^2, \varphi^\beta] = \E^{\I S^\beta_\mathrm{IF} [\varphi^1, \varphi^2, \varphi^\beta, t]} = \\
        &\int_{\mathscr{C}_\beta} \mathcal{D} \phi \; \exp \Bigg[ \I S_2 [\phi] + \I \int_{t,\partial \Omega} \left( \frac{\partial \phi^2}{\partial n} \varphi^2 - \frac{\partial \phi^1}{\partial n} \varphi^1 \right) \\
        &\hspace*{2em} - \int_{\beta,\partial \Omega} \frac{\partial \phi^\beta}{\partial n} \varphi^\beta \Bigg] \; ,
    \end{split}
\end{equation}
where we used \eqref{eq:S_12} and interpreted $\phi$ as a Dirichlet field in the exterior region. Written like this, we may interpret the influence functional $\mathfrak{F}^\beta_\mathrm{IF}$ as the generating functional of path ordered correlation functions of normal derivatives of the field on the boundary for the exterior Dirichlet theory, where the interior field $\varphi$ acts as a (time and path dependent) source term on the boundary. Similarly, the influence action $S^\beta_\mathrm{IF}$ can be interpreted as the generating functional of connected path ordered correlation functions of normal derivatives of the field on the boundary.

Notice that for vanishing boundary source fields, the expectation value of the normal derivative on the boundary of $\phi$ is zero, i.e., $\braket{\partial \phi (\mathrm{x}) / \partial n_x}_\beta |_{\varphi^1 = \varphi^2 = \varphi^\beta = 0} = 0$ for all $\mathrm{x} \in [0,t] \times \partial \Omega$. However, in the presence of boundary sources, these expectation values are given by
\begin{equation}\label{eq:boundary_source_expectation_value}
    \begin{split}
        &\left\langle \frac{\partial \phi^1 (\mathrm{x})}{\partial n} \right\rangle_{\beta, \varphi^1, \varphi^2, \varphi^\beta} = \frac{\delta}{\delta \varphi^1 (\mathrm{x})} S^\beta_\mathrm{IF} [\varphi^1, \varphi^2, \varphi^\beta] \\
        &= \I \int_{t,\partial \Omega} \left( g^{11} (\mathrm{x}, \mathrm{y}) \varphi^1 (\mathrm{y}) - g^{12} (\mathrm{x}, \mathrm{y}) \varphi^2 (\mathrm{y}) \right) \\
        &- \int_{\beta, \partial \Omega} g^{1 \beta} (\mathrm{x}, \mathrm{y}) \varphi^\beta (\mathrm{y})  \; ,
    \end{split}
\end{equation}
and analoguesly for $\partial \phi^2 / \partial n$ unpon the interchange $1 \leftrightarrow 2$. The sources in the above expression are still fluctuating. Upon averaging over the interior field $\varphi$ and denoting its expectation value again by $\Phi$, \eqref{eq:boundary_source_expectation_value} reads
\begin{equation}
    \left\langle \frac{\partial \phi^i (\mathrm{x})}{\partial n_x} \right\rangle_{\beta, \Phi} =\I \int_{t,\partial \Omega} \left( g^{ii} (\mathrm{x}, \mathrm{y}) - g^{ij} (\mathrm{x}, \mathrm{y}) \right) \Phi (\mathrm{y}) \; ,
\end{equation}
where we used the fact that the expectation value of $\varphi$ on the boundary $\partial \Omega$ vanishes on the imaginary time branch. Using this result, we see that the expectation value of the normal derivative of the exterior difference field $\Upsilon \coloneqq \phi^1 - \phi^2$ on the boundary is given by
\begin{equation}
    \left\langle \frac{\partial \Upsilon (\mathrm{x})}{\partial n_x} \right\rangle_{\beta, \Phi} = \int_{t,\partial \Omega} \boldsymbol{\mathrm{D}} (\mathrm{x},\mathrm{y}) \, \Phi (\mathrm{y}) \; ,
\end{equation}
where we used the definition of the dissipation kernel $\boldsymbol{\mathrm{D}}$ in \eqref{eq:dissipation_kernel}.

Therefore, we see that the dissipation kernel $\boldsymbol{\mathrm{D}}$ describes the response of the normal derivative of the field on the boundary to the perturbation given by the linear coupling of the normal derivative to the classical ``force'' $\Phi$. As expected the \emph{response function} $\boldsymbol{\mathrm{D}}$ is retarded (i.e., causal), since the force $\Phi$ cannot cause an effect before the force is applied \cite{Altland2010}.

\subsubsection{Energy Non-Conserving Boundary Conditions}

Finally, we show that the boundary condition \eqref{eq:stochastic_boundary_condition} of the equation of motion for the field expectation value in the interior region \eqref{eq:stochastic_Klein_Gordon} does not conserve energy due to the presence of the dissipation kernel $\boldsymbol{\mathrm{D}}$. More precisely, we say that our equation of motion is energy non-conserving if a suitably defined energy functional is not constant in time. If the change of energy is non-positive, we say that the equation of motion is dissipative\footnote{For an overview of dissipative operators on Hilbert spaces and the closely related concept of accretive operators, we refer to, e.g., \cite{Exner1985,Kato1995}. Dissipative hyperbolic systems of partial differential equations with particular emphasis on dissipative boundary conditions are treated in, e.g., \cite{Phillips1957,Friedrichs1958,Phillips1959,Lax1960,Lagnese1983,Triggiani1989,Bielak1990,Messaoudi2010,Petkov2016,Eller2022}. The following discussion is inspired by the energy methods known from the theory of partial differential equations, see, e.g., \cite[Sec.~2.4.3]{Evans2010}, where this concept is demonstrated on the example of the wave equation.}.

We define the energy form $\mathfrak{E}_t [\Phi]$ of a solution of the Klein-Gordon initial-boundary value problem \eqref{eq:stochastic_Klein_Gordon} -- \eqref{eq:initial_conditions} at time $t \geq 0$ as
\begin{equation}
    \mathfrak{E}_t [\Phi] = \frac{1}{2} \int_\Omega \left( |\dot{\Phi}|^2 + \left| \nabla \Phi \right|^2 + m^2 |\Phi|^2 \right) \D^d x \; ,
\end{equation}
where the field $\Phi$ on the right-hand side is taken at fixed time $t$. Since the energy of a complex field $\Phi$ is just the sum of the energies of its real and imaginary parts, we will from now on restrict the discussion to real solutions $\Phi$. The change of energy is then given by
\begin{equation}
    \begin{split}
        \frac{\D}{\D t} \mathfrak{E}_t [\Phi] &= \int_\Omega \left( \dot{\Phi} \ddot{\Phi} + \nabla \Phi \cdot \nabla \dot{\Phi} + m^2 \Phi \dot{\Phi} \right) \D^d x \\
        &= \int_{\partial \Omega} \dot{\Phi} \frac{\partial \Phi}{\partial n} \; \D S \; ,
    \end{split}
\end{equation}
where in the second step we used Green's first identity and the fact that $\Phi$ satisfies the Klein-Gordon equation \eqref{eq:stochastic_Klein_Gordon}. We see that the change of energy is exclusively due to a boundary term and no bulk sources or sinks are present. Furthermore, for manifestly energy conserving boundary conditions such as Dirichlet ($\dot{\Phi} |_{\partial \Omega} = 0$) or Neumann ($\partial \Phi / \partial n = 0$) boundary conditions, the change of energy vanishes.

Under the assumption that $\Phi$ is a solution of the system \eqref{eq:stochastic_Klein_Gordon} -- \eqref{eq:initial_conditions}, we can write the change of energy of such a solution as
\begin{equation}
    \frac{\D}{\D t} \mathfrak{E}_t [\Phi] = - \int_{\partial \Omega} \dot{\Phi} \left( \hat{\boldsymbol{\mathrm{D}}} \Phi + \xi \right) \D S \; .
\end{equation}
Since $\xi$ is a centred Gaussian random variable, we can ignore it when considering averages. Therefore, the change of energy is purely due to the presence of the dissipation kernel $\boldsymbol{\mathrm{D}}$ in the boundary condition \eqref{eq:stochastic_boundary_condition}.

\subsubsection{One-Dimensional Wave Equation}

As a final consistency check, we consider the hyperbolic system \eqref{eq:stochastic_Klein_Gordon} -- \eqref{eq:initial_conditions} for $d = 1+1$, $\Omega = (0,\infty)$ and $m = 0$, i.e., we consider the one-dimensional wave equation on the positive half line. In this case, the retarded Green's function for the exterior region with Dirichlet boundary conditions on the spatial boundary $\partial \Omega = \{0\}$ is given by
\begin{equation}
    \begin{split}
        &G^{\mathrm{ret}}_\mathrm{D} (x,\tau;y,\tau') = \\
        &\frac{1}{2} \left( \Theta (\tau - \tau' - |x+y|) - \Theta (\tau - \tau' - |x-y|) \right) \; ,
    \end{split}
\end{equation}
where $x, y \in (-\infty,0]$.

The dissipation kernel $\boldsymbol{\mathrm{D}}$ is the double normal derivative of this exterior retarded Green's function on the boundary $\partial \Omega = \{0\}$, which evaluates to
\begin{equation}
    \begin{split}
        &\boldsymbol{\mathrm{D}} (\tau - \tau') \\
        &= \left. \frac{\partial^2}{\partial x \partial y} G^{\mathrm{ret}}_\mathrm{D} (x,\tau;y,\tau') \right|_{x=y=0} = \left. \frac{\partial}{\partial s} \delta (s) \right|_{s=\tau - \tau'} \; .
    \end{split}
\end{equation}
Therefore, the operator $\hat{\boldsymbol{\mathrm{D}}}$ acts on a solution $\Phi$ of the wave equation as
\begin{equation}
    (\hat{\boldsymbol{\mathrm{D}}} \Phi) (t) = \dot{\Phi} (t,0) \; .
\end{equation}
The boundary condition \eqref{eq:stochastic_boundary_condition} then takes the form
\begin{equation}
    \frac{\partial \Phi}{\partial n} (t,0) + \dot{\Phi} (t,0) = - \xi (t,0) \; .
\end{equation}
Except for the stochastic force term $- \xi$, the above expression describes \emph{transparent} (or \emph{non-reflecting}) boundary conditions for the one-dimensional wave equation, see, e.g., \cite[Sec.~2.1]{Ionescu2003}.

The change of energy of a complex solution $\Phi$ for the above transparent boundary conditions reads
\begin{equation}
    \frac{\D}{\D t} \mathfrak{E}_t [\Phi] = - | \dot{\Phi} |^2 \; ,
\end{equation}
where we averaged out the stochastic term $- \xi$. We see that for transparent boundary conditions, the change of energy of the system is non-positive, and the system is therefore fully dissipative.

\section{Interacting Theories}\label{sec:Interacting_Theories}

In this Section, we discuss the generalization of the results obtained in the previous Sections to interacting theories. More precisely, we consider the case of a (lattice regularized) massive scalar field theory with polynomial self-interaction, i.e., we assume that $V$ in \eqref{eq:Hamiltonian_operator_anharmonic} is a polynomial bounded from below. For the initial state we will again consider the case of a local perturbation of a thermal state as discussed in Section \ref{sec:Beyond_the_Product_Ansatz} for the non-interacting theory. We will show that the non-unitary time evolution of the reduced state is still encoded in an effective boundary action.

For this model, the classical and Euclidean actions $S$ and $S_\mathrm{E}$ split into interior, exterior and boundary components as
\begin{align}
    S (\varphi, \phi) &= S_1 (\varphi) + S_2 (\phi) + S_{12} (\varphi, \phi) \; , \\
    S_\mathrm{E} (\varphi, \phi) &= S_\mathrm{E}^1 (\varphi) + S_\mathrm{E}^2 (\phi) + S_\mathrm{E}^{12} (\varphi, \phi) \; ,
\end{align}
where $S_1$, $S_\mathrm{E}^1$ and $S_2$, $S_\mathrm{E}^2$ are the action functionals associated with the regions $\Lambda_1$ and $\Lambda_2$, respectively, given by
\begin{align}
    \begin{split}
        &S_i (\psi) = \\
        &\frac{1}{2} \int_{0}^{t} \D \tau \int_{\Lambda_i} \D^d x \left[ \dot{\psi}_x^2 - \int_{\Lambda_i} \D^d y \; \psi_x \, D^{\varepsilon,i}_{xy} \, \psi_y - V (\psi) \right] \; ,
    \end{split} \\
    \begin{split}
        &S^i_\mathrm{E} (\psi) = \\
        &\frac{1}{2} \int_0^\beta \D \tau \int_{\Lambda_i} \D^d x \left[ \dot{\psi}_x^2 + \int_{\Lambda_i} \D^d y \; \psi_x \, D^{\varepsilon,i}_{xy} \, \psi_y + V (\psi) \right] \, ,
    \end{split}
\end{align}
for $i \in \{ 1,2 \}$ and $S_{12}$ and $S_\mathrm{E}^{12}$ couple the fields at the boundary, i.e.,
\begin{align}
    S_{12} (\varphi, \phi) &= - \int_{0}^{t} \D \tau \int_{\partial \Lambda} \D S (x) \; (\partial^\varepsilon_n \phi_x) \, \varphi_x \; , \\
    S_\mathrm{E}^{12} (\varphi, \phi) &= - \int_0^\beta \D \tau \int_{\partial \Lambda} \D S (x) \; (\partial^\varepsilon_n \phi_x) \, \varphi_x \; ,
\end{align}
where we again interpret $\phi$ as a Dirichlet field in the exterior region in order to write the boundary actions in terms of (discrete) normal derivatives. We observe that since the potential $V$ does not couple neighbouring lattice sites, the boundary actions $S_{12}$ and $S_\mathrm{E}^{12}$ are the same as in the non-interacting case. In particular, we can interpret them as source terms on the spatial boundary.

With this setup, the influence functional for the interacting theory is given by
\begin{equation}
    \begin{split}
        &\mathfrak{F}^\beta_\mathrm{IF} [\varphi^1, \varphi^2, \varphi^\beta] = \E^{\I S^\beta_\mathrm{IF} [\varphi^1, \varphi^2, \varphi^\beta, t]} = \\
        &\int_{\mathscr{C}_\beta} \mathcal{D} \phi \; \exp \left[ \I S_2 [\phi] - \I \int_{\mathscr{C}_\beta ,\partial \Omega} \frac{\partial \phi}{\partial n_x} \varphi \right] \; .
    \end{split}
\end{equation}
Once again, the influence functional is the generating functional of path ordered correlation functions of the normal derivative of the field on the boundary for the exterior Dirichlet theory, where the interior field $\varphi$ acts as a (time and path dependent) source term on the boundary. Similarly, the influence action $S^\beta_\mathrm{IF}$ can be interpreted as the generating functional of connected path ordered correlation functions of normal derivatives of the field on the boundary in the presence of the boundary source $\varphi$.

\section{Conclusion}\label{sec:Conclusion}

In this work, we have studied the time evolution of states of relativistic quantum field theories restricted to a spatial subregion. We have shown that the time evolution is non-unitary and that the reduced state evolves in time like an open quantum system. In particular, the dynamical coupling between the system (the degrees of freedom within some spatial region) and the environment (the degrees of freedom in the complement of that region) occurs via the differential operator in the classical action. Due to this local structure, the coupling between the system and the environment is linear and only across the boundary of the region. Therefore, the effects of the system-environment coupling on the time evolution of the reduced density operator are entirely contained in an effective boundary action. We note that this is true for both non-interacting and interacting theories as long as the interactions are ``ultra-local'', i.e., they do not couple neighbouring lattice sites.

In addition to the linear system-environment coupling, it is necessary to consider initial state correlations. This is due to the fact that states in relativistic quantum field theories are generically entangled across spacetime regions. We incorporated initial state correlations by considering (global) thermal states with local excitations. Since the excitations are purely contained within the interior region, the initial state correlations are only due to the global thermal state, which can be taken care of using a Euclidean path integral representation. It is noteworthy that the incorporation of initial state correlations results in the emergence of an effective boundary term in the action, exhibiting a structural similarity to the one induced by the system-environment coupling. In particular, no bulk terms are induced by the initial state correlations considered here.

It is worth noting that the boundary terms encoding the non-unitary contributions to the time evolution of the reduced state are similar to the free boundary conditions considered, e.g., in \cite{Guerra1975a,Guerra1975b,Guerra1976,Floerchinger2023}. More specifically, the dissipation and noise kernels in the non-interacting model are both given by a double normal derivative of an exterior Dirichlet $2$-point function. In particular, in Section \ref{sec:Euclidean_Field_Theory} we considered a Euclidean field theory and demonstrated that the splitting of the action yields a boundary action that precisely describes the free boundary conditions already encountered in the literature. This indicates that the aforementioned structure of the boundary action, namely a double normal derivative of some exterior Dirichlet Green's function, is a generic phenomenon when considering the restriction of thermal states to spatial subregions.

Finally, we derived a stochastic equation of motion for the field expectation value in a spatial region for the free theory. Structurally, this equation is just the usual Klein-Gordon equation but with stochastic spatial boundary conditions. More precisely, these boundary conditions are induced by the dissipation and noise kernels previously shown to be entirely supported on the boundary. We argue that this partial differential equation is interesting on its own right, especially concerning the boundary conditions which are non-local in both space and time. It would be of interest to investigate the properties of solutions to such an equation, in particular the emergence of non-trivial boundary effects due to the stochastic boundary conditions.

We now propose a number of avenues for further investigation. First, it would be beneficial to investigate whether and in what sense the non-unitary time evolution of the reduced state causes local thermalization and thus a first-principle explanation of the emergence of an effective fluid dynamic description of quantum field theories. It is clear that in order to truly observe thermalization, one must consider interacting theories\footnote{Note, however, the concept of generalized thermalization in isolated integrable systems \cite{Vidmar2016,Calabrese2020}.}. In order to do so, the discussion teased in Section \ref{sec:Interacting_Theories} needs to be developed further using methods from non-equilibrium quantum field theory, like, e.g., the $n$-PI formalism \cite{Calzetta2022,Berges2004,Berges2015}.

Secondly, it is necessary to define precisely what is meant by local thermalization, which may be an information-theoretic question. As argued in, for example, \cite{Dowling2020} and \cite{Mueller2015}, a reasonable notion of subsystem thermalization is the vanishing of the quantum \emph{relative} entropy $S (\hat{\rho}_\Lambda \| \hat{\sigma}^\beta_\Lambda)$ \cite{Umegaki1962,Vedral2002,Nielsen2012}, where $\hat{\rho}_\Lambda$ is the reduced state of a subsystem $\Lambda$ and $\hat{\sigma}^\beta_\Lambda$ is the \emph{reduced} thermal state of the same subsystem of some inverse temperature $\beta$. The use of relative entropies is motivated by the fact that relatives entropies are well-defined also for systems with infinitely many degrees of freedom. For examples of the use of relative entropy in quantum and statistical field theories, see \cite{Casini2008,Blanco2013,Witten2018,Hollands2018,Casini2019,Longo2019,Floerchinger2022,Floerchinger2023,Ditsch2023}. It would be of interest to ascertain whether this concept of thermalization can be related to the local time evolution of the reduced state within the context of the model presented here.

Finally, we note that the concepts derived in this work can be used to study the time evolution of entanglement entropy in quantum field theories. For example, in the case of a purely Gaussian state in a lattice regularized scalar field theory, we can use results from continuous variable quantum information theory to calculate the entanglement entropy as a function of time \cite{Serafini2017}. More precisely, using Williamson's Theorem \cite{Ikramov2018}, it is sufficient to compute the symplectic eigenvalues of the covariance matrix of the reduced (Gaussian) state of the interior region to determine the entanglement entropy. The time dependence of the symplectic spectrum can then be studied using the techniques developed in this paper. While closed-form expressions for the entanglement entropy are not expected in general, we expect that the entanglement entropy as a function of time can be evaluated numerically for sufficiently few lattice sites. For an example of how Williamson's theorem is used to calculate entanglement entropy in a quantum field theory, see e.g. \cite{Berges2018b}.

% Specify following sections are appendices. Use \appendix* if there
% only one appendix.
% \appendix*

% If you have acknowledgments, this puts in the proper section head.
\begin{acknowledgments}
    We acknowledge valuable discussions with Christian Schmidt and Tim St\"otzel and thank Tim St\"otzel for a careful reading of the manuscript. This work is supported by the Deutsche Forschungsgemeinschaft (DFG, German Research Foundation) under 273811115 -- SFB 1225 ISOQUANT.
\end{acknowledgments}

% Create the reference section using BibTeX:
\bibliography{references}

\end{document}